\documentclass[3p]{elsarticle}
\def\Section {\S}
\usepackage[numbers]{natbib}

\def\tsc#1{\csdef{#1}{\textsc{\lowercase{#1}}\xspace}}
\tsc{WGM}
\tsc{QE}

\AtBeginDocument{%
  \providecommand\BibTeX{{%
    \normalfont B\kern-0.5em{\scshape i\kern-0.25em b}\kern-0.8em\TeX}}}

\usepackage{graphicx}
\usepackage{balance,url,textcomp,amsmath,multirow}
\usepackage{tabularx,ragged2e,booktabs,caption}
\usepackage[linesnumbered,ruled]{algorithm2e}
\usepackage{graphicx}
\usepackage{subfigure}
\usepackage{multirow}
\usepackage[table,xcdraw]{xcolor}
\usepackage{xfrac}    
\usepackage{geometry}
\usepackage{amsmath}
\usepackage{siunitx}
\DeclareSIUnit\E{E}
\DeclareSIUnit\c{c}
\newcommand{\systemName}{\texttt{WattScope}\xspace}

\usepackage{amssymb}

\journal{Performance Evaluation}

\begin{document}

\begin{frontmatter}


\title{WattScope: Non-intrusive Application-level Power Disaggregation\\ in Datacenters}


\author{Xiaoding Guan}
\author{Noman Bashir}
\author{David Irwin}
\author{Prashant Shenoy}
\address{University of Massachusetts Amherst}

\begin{abstract}
Datacenter capacity is growing exponentially to satisfy the increasing demand for many emerging computationally-intensive applications, such as deep learning.  This trend has led to concerns over datacenters' increasing energy consumption and carbon footprint.  The most basic prerequisite for optimizing a datacenter's energy- and carbon-efficiency is accurately monitoring and attributing energy consumption to specific users and applications. Since datacenter servers tend to be multi-tenant, i.e., they host many applications, server- and rack-level power monitoring alone does not provide insight into the energy usage and carbon emissions of their resident applications.  At the same time, current application-level energy monitoring and attribution techniques are \emph{intrusive}: they require privileged access to servers and necessitate coordinated support in hardware and software, neither of which is always possible in cloud environments.  To address the problem, we design \systemName, a system for non-intrusively estimating  the power consumption of individual applications using external measurements of a server's aggregate power usage and without requiring direct access to the server's operating system or applications.  Our key insight is that, based on an analysis of production traces, the power characteristics of datacenter workloads, e.g., low variability, low magnitude, and high periodicity, are highly amenable to disaggregation of a  server's total power consumption into application-specific values.  \systemName adapts and extends a machine learning-based technique for disaggregating building power and applies it to server- and rack-level power meter measurements that are already available in data centers. We evaluate \systemName's accuracy on a production workload and show that it yields high accuracy, e.g., often $<$$\sim$10\% normalized mean absolute error, and is thus a potentially useful tool for datacenters in externally monitoring application-level power usage.
\end{abstract}

\end{frontmatter}

\section{Introduction}
\label{sec:introduction}
Datacenter capacity is growing exponentially to satisfy the increasing demand for many emerging computationally intensive applications.  For example, a recent analysis estimated a 6$\times$ increase in datacenter capacity from 2010-2018 (or $\sim$22\% per year)~\cite{masanet} with capacity doubling in the past five years~\cite{srgresearch20}.   This capacity increase is being driven by a variety of emerging application classes, such as cryptomining~\cite{bitcoin-energy}, machine learning (ML), and other big-data processing. As one example, over the past decade, the cycles devoted to training state-of-the-art ML models has been doubling every $3.4$ months, which is much faster than Moore's Law~\cite{ml-compute-demand}.  Of course, increases in datacenter capacity have also led to increases in their energy consumption despite substantial improvements in their energy-efficiency over the past decade~\cite{growth,hintemann2020efficiency, andrae2019projecting, andrae2015global, belkhir2018assessing}.  Unfortunately, datacenter energy usage is poised to increase substantially in the coming decade due to the end of Dennard scaling and limited opportunities for further significant improvements in datacenter energy efficiency. For example, Google datacenters' Power Usage Effectiveness (PUE)---the ratio of their total energy to the energy of IT equipment---is now $\sim$$1.1$, which is already near the optimal value of $1$~\cite{google-pue}.  

The trends above have led to increasing concern and criticism over datacenters' energy consumption and their resulting carbon footprint.   As a result, many cloud providers and datacenter operators have begun to increase their emphasis on energy-efficient and sustainable operations.  Indeed, prominent technology companies, including Google, Amazon, Meta, and Microsoft, have set ambitious goals to become carbon-neutral~\cite{amazon-carbon-neutral,facebook-carbon-neutral,vmware-carbon}, carbon-free~\cite{google-carbon-free}, or even carbon-negative~\cite{microsoft-carbon-negative} within the next 10-20 years. Importantly, \emph{the simplest and most basic prerequisite for optimizing a datacenter's energy- and carbon-efficiency is providing applications visibility into their power consumption, as they cannot optimize a metric they cannot measure}. Datacenters are well-instrumented with external power meters typically attached to rack-level power distribution units (PDUs) and individual servers. However, rack- and server-level power monitoring does not provide insight into the power consumed by individual applications, since servers are multi-tenant and host multiple applications. Even when a server runs a single application, external power usage data is often not exposed to the application.  While some servers may support internal hardware-level power monitoring, such as Intel's RAPL~\cite{rapl}, they cannot directly monitor power at the granularity of individual applications. In addition, internal hardware-level power monitoring is typically a highly inaccurate measurement of the total system power (with up to 70\% error), as we discuss \S\ref{sec:background}, since it only measures the power of CPU sockets and memory, and thus does not capture the power usage of any other important system components, such as the power supply, motherboard, I/O devices, and GPUs.  Notably, these other system components are accounting for an increasingly large share of server power consumption. 

Prior work has developed techniques for attributing server-level power to applications, which generally involve 
apportioning a server's total power usage based on each application's resource utilization. One common approach involves training a model that takes per-process hardware performance counters as input and infers a corresponding power usage, e.g., from RAPL.  For example, PowerAPI is an open-source toolkit that uses such techniques to monitor application-level power~\cite{powerapi,powerapi2,powerapi3}. However, such approaches require access to the hardware performance counters. There are many scenarios, which we summarize below, where access to hardware counters is either not available or too intrusive.  (i) Most importantly,  prior approaches do not apply well to cloud users. While cloud providers can use prior approaches to track the power consumption of users' virtual machines, users that host multiple applications within each virtual machine cannot attribute power to each application, since they lack privileged access to hardware counters. Thus, existing techniques are not applicable to cloud users. (ii) In addition, process-level power monitoring techniques are intrusive, as they require server resources that scale with the number of processes tracked, as well as the power data resolution. This overhead can be high when tracking many applications at high resolution, and becomes prohibitive at some point. (iii) Further, since hardware interfaces are not standardized, existing \emph{in situ} techniques are not general and must be tailored to specific hardware platforms.  For example, PowerAPI uses RAPL, which is only available on Intel processors. Since the set of hardware counters also varies by platform, existing power monitoring toolkits are primarily designed for Intel or AMD platforms, but generally do not support Power-, ARM-, and RISCV-based platforms.  The limitations above are the primary reason that fine-grained application-level power monitoring is not offered by cloud providers.  This lack of support in-turn prevents cloud applications from optimizing their energy consumption and carbon emissions. Ultimately, the lack of application-level visibility into energy consumption is a key impediment to achieving the ambitious sustainability goals above, as it is impossible to optimize a metric that cannot be effectively measured.

To address the problem, we design \systemName, a system for non-intrusively monitoring application-level power consumption using aggregate server-level power measurements. \systemName uses disaggregation techniques to 
apportion power data from external server- and rack-level power meters, which are typically available in power distribution units (PDUs), into individual application-level power usage without requiring intrusive access to system and application software.  \systemName recognizes that datacenters already collect server- and rack-level power data for thermal management and billing purposes, which can be leveraged to also provide application-level power monitoring. Thus, \systemName analyzes power data collected from these external meters to infer each application's power usage.  More formally, \systemName disaggregates a time-series of power readings $P(t)$, over some sampling interval $\Delta$$t$, into a separate time-series $p_i(t)$ for each individual application $i$, such that $\forall t, P(t) = \sum_i p_i(t)$. Importantly, \systemName does not require any server-level access or specific hardware/software support, and instead can run externally as part of the facility management system.  As a result, \systemName can be deployed in nearly any datacenter facility with PDUs that measure server- and rack-level power. Our key insight is that, based on a large-scale analysis of production traces, the power characteristics of datacenter workloads, e.g., low variability, low magnitude, and high periodicity, are highly amenable to disaggregation.  \systemName adapts and extends a deep learning-based technique, originally designed for disaggregating building power, and applies it to servers and racks.  We implement \systemName and experimentally evaluate its accuracy on a production workload. 

Our hypothesis is that disaggregating server- and rack-level power using \systemName can enable highly accurate and non-intrusive application-level power monitoring without requiring any server-level hardware/software support.  In evaluating our hypothesis, we make the following contributions. 

\noindent {\bf Production Workload Analysis}.  We first analyze the job characteristics of a large-scale production workload from a major cloud provider that includes 5-minute resource usage readings for $2.7$ million jobs over a 30 day period encompassing more than $100$ million job-hours. Our analysis reveals that job usage patterns exhibit multiple characteristics, including low variability, low magnitude, and high periodicity, that \systemName can potentially exploit for disaggregation. Our analysis also shows that, while server applications can operate arbitrarily and irregularly in general, they have a high degree of regularity in practice.

\noindent {\bf WattScope Design}.  We present \systemName's design, which adapts and extends a deep learning-based disaggregation algorithm originally applied to building power data.  \systemName's design includes a library of models trained for different classes of applications based on their variability, magnitude, and periodicity.  \systemName then integrates with a cluster scheduler to learn the number and type of applications running on each server, i.e., based on their attributes, to select an appropriate model for disaggregation.  

\noindent {\bf Implementation and Evaluation}.  Finally, we implement and evaluate a \systemName prototype.  We implement \systemName's disaggregation technique by modifying nilmtk-contrib~\cite{nilmtk-contrib}, an open-source reference implementation of multiple algorithms for building energy disaggregation, to instead disaggregate server- and rack-level power, and evaluate accuracy across multiple dimensions using our production workload trace.  We evaluate \systemName's accuracy on a production workload and show that it yields high accuracy, e.g., often $<$$\sim$10\% normalized mean absolute error, and is thus a potentially useful tool for broadly enabling application-level power monitoring in datacenters. 

\section{Motivation and Background}
\label{sec:background}
A key motivation for our work is that providing application-level visibility and monitoring of power usage is essential to satisfying companies' ambitious sustainability goals. Indeed, the U.S. Securities and Exchange commission may soon require companies, including those using shared server/cloud resources, to report their carbon emissions from energy usage~\cite{sec-rule}. In addition, while there has long been a strong incentive to optimize computing's energy-efficiency, since energy incurs a cost, optimizing computing's carbon emissions is different, as energy's carbon-intensity varies over time and by region based on the energy mix, e.g., fossil fuels, nuclear, and renewables~\cite{war}.  As a result, reducing carbon emissions often necessitates monitoring and adapting application power usage over time in response to changes in energy's carbon-intensity.

Our work assumes that a datacenter has external power meters deployed at each server (or rack) that are capable of continuously monitoring server (or rack) power $P(t)$ (in watts) over some interval $\Delta$$t$, e.g., every 5 minutes. Datacenter servers generally include external power meters and make them available programmatically in real-time via network protocols, such as IPMI~\cite{ipmi} or Redfish~\cite{redfish}, to facility management systems.  External power monitoring is necessary for basic datacenter operations, such as fault identification and billing.  In addition, even if individual servers do not include external power meters, power distribution units (PDUs) that provide power to servers in one or more racks also typically include them.  

\begin{figure}[t]
\centering
\begin{tabular}{@{}c@{}c}
\includegraphics[width=0.5\linewidth]{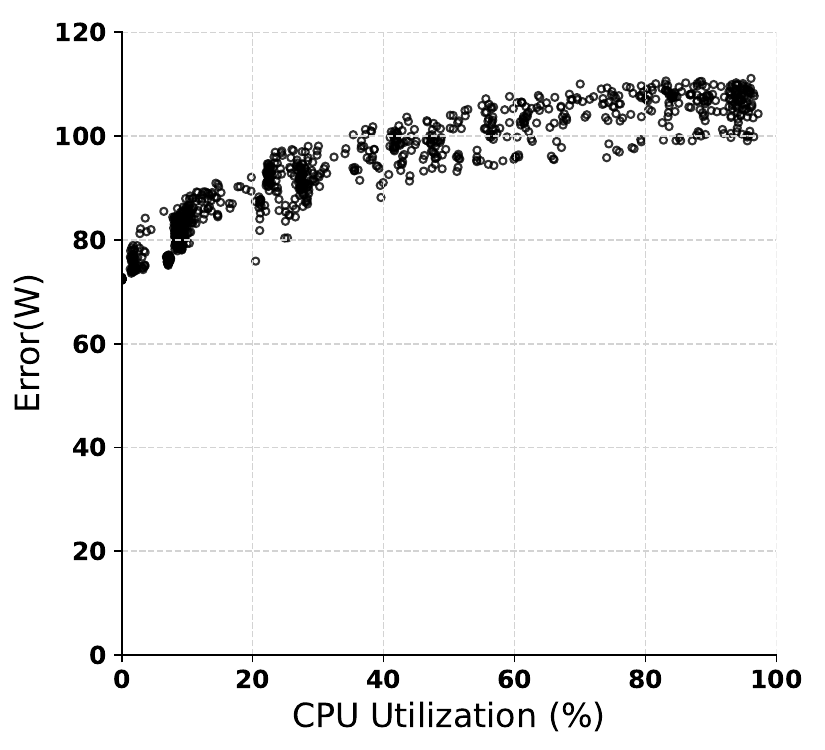} &
\includegraphics[width=0.5\linewidth]{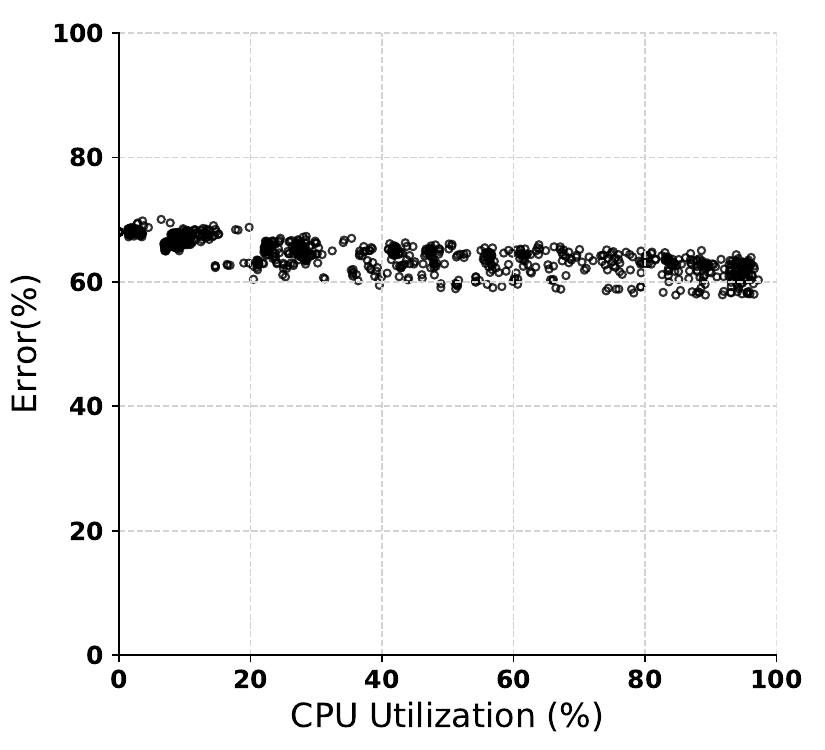}\\
(a) Absolute Error & (b) Percentage Error
\end{tabular}
\vspace{-0.25cm}
\caption{\emph{\textbf{Absolute (a) and percentage (b) error in measuring a traditional server's power using RAPL, which is Intel's internal power monitoring platform.  
}}}
\label{fig:rapl-example}
\vspace{-0.5cm}
\end{figure}

In general, the data above is collected as part of facility management and is not exposed to application- or system-level software. Instead, application- and system-level power monitoring typically uses internal hardware and software support.  For example, PowerAPI~\cite{powerapi,powerapi2,powerapi3} leverages a model that maps hardware counters and RAPL readings to application-level power consumption. However, since hardware support is not standardized, this approach is not general.  In addition, as mentioned in \Section\ref{sec:introduction}, RAPL, which measures CPU socket and memory power, often does not provide an accurate measurement of full system power.  To illustrate Figure~\ref{fig:rapl-example} shows the absolute (a) and percentage (b) error in RAPL power measurements for a traditional server compared to an external power meter that directly measures the server's power. In this case, the server's maximum power at 100\% utilization is 175W.   The figure illustrates that since RAPL measurements only account for a subset of the server's hardware resources, they capture only between 30-40\% of the total power a server actually consumes.  In addition, RAPL measurement errors vary widely---from 75-110W (a) or 60-70\% (b)---depending on the server's utilization.  In addition, these errors would likely be much worse for modern servers with GPUs, since RAPL measurements do not include GPUs. While GPUs often have their own internal interfaces for monitoring power, these are also not standardized or general.  Of course, RAPL only measures CPU socket and memory power, and \emph{not} application power, so even if RAPL measurements were accurate, an additional server-specific model is necessary to map application resource usage, e.g., using hardware counters, to the fraction of power an application consumes.

Importantly, \systemName \emph{does not} assume any hardware or software support on any server, and does not require training a server-specific model.  However, after disaggregating a server's power into the power usage of individual applications, mapping the applications to specific application names running on servers is often important. Thus, \systemName assumes a minimal interface to integrate with a cluster-level scheduler that provides, in addition to the number of applications running on a server, the names of the applications running on it and a minimal amount of coarse summary usage characteristics, e.g., variability, magnitude, and periodicity. As we discuss, schedulers typically provide the former, while the latter is simple to implement.  

\systemName builds on prior work on  energy disaggregation or non-intrusive load monitoring (NILM) for buildings, which infers the average power usage $p_i(t)$ of each building load $i$ at time $t$ given the building's average power usage $P(t)$ measured at an external smart meter over some interval $\Delta$$t$.  Importantly, the prior work on NILM has shown that disaggregation accuracy varies widely based on loads' power signatures, i.e., their pattern of power usage~\cite{nilmtk-contrib}. For example, disaggregating large loads, such as electric dryers, is more accurate than small ones, since the power signature of such loads is more distinctive in a building's aggregate power usage.  Similarly, disaggregating highly periodic loads and those with discrete power states, such as refrigerators or water heaters, is more accurate than noisy loads that are highly variable, such as many electronics. Finally, larger buildings with more loads decrease disaggregation accuracy, as it becomes more difficult for algorithms to extract power signatures for individual loads.  

Given the importance of power usage characteristics on the effectiveness of disaggregation, we next analyze a production workload trace to understand the resource and power usage characteristics of real cloud applications.  Since server applications can exhibit highly variable resource and power usage, there is no guarantee that their characteristics will be amenable to disaggregation, as with electric dryers, water heaters, refrigerators, etc. In addition, unlike buildings, which consist of a common set of appliances, the number and types
of server applications is not fixed.  As a result, it is not clear a priori whether disaggregation methods can be applied to server applications.

\section{Production Workload Analysis } 
\label{sec:workload}

To guide \systemName's design, we conduct a large-scale analysis of production workloads, that provide information for individual virtual machines (VMs), containers, or application tasks, to quantify their regularity, variability, and intensity i.e., magnitude, in resource and power usage.  

\subsection{Analysis Setup}
\label{sec:analysis-setup}
Below, we provide details on the workload traces, power models, and metrics we use in our analysis.

\vspace{0.1cm}
\textbf{Workload Traces.} To evaluate \systemName's efficacy, we require a dataset that provides ground truth power information for different VMs or processes running on a server along with the server's aggregate power usage.  While external power meters can record server-level power consumption, it is not possible to instrument individual VMs or processes with a physical power meter and record their usage, as they are virtual and not physical.  As discussed in \S\ref{sec:introduction}, prior work has developed other methods using hardware performance counters to build models that estimate per-VM or per-process power consumption~\cite{power-api}. However, such methods are highly intrusive, incur an overhead, and are thus not widely deployed in practice. As a result, to the best of our knowledge, there is no publicly available dataset that provides power usage information for individual VMs and application processes on servers.  Consequently, we construct such a ground-truth trace from publicly-available CPU and memory workload traces and use them to derive server power consumption, i.e., by mapping the usage information to power.

To generate power consumption traces for our analysis, and later evaluation of \systemName in \S\ref{sec:evaluation}, we use two of the most commonly used industry workload traces: Microsoft Azure Traces (V2)~\cite{azure-trace} and Google Cluster Workload Traces (V3)~\cite{google-trace}. The Azure trace provides the minimum, maximum, and average CPU utilization and memory allocation for $\sim$2.7 million production VMs every 5 minutes over a 30-day period.  The Azure trace has a size of 235GB and contains $\sim$1.9 billion readings. The Google cluster workload trace provides average CPU usage, CPU usage histograms, and memory usage information of jobs for each 5 minute period over a 31-day period.  The Google trace contains data for $\sim$2.5 million jobs from 96.4k physical servers spread across 8 datacenters. In the Azure dataset, we assume each VM hosts a different application or job. For uniformity and ease of exposition, henceforth, we also refer to Azure VMs  as jobs. 

\begin{figure}[t]
\centering
\begin{tabular}{@{}c@{}c}
\includegraphics[width=0.5\linewidth]{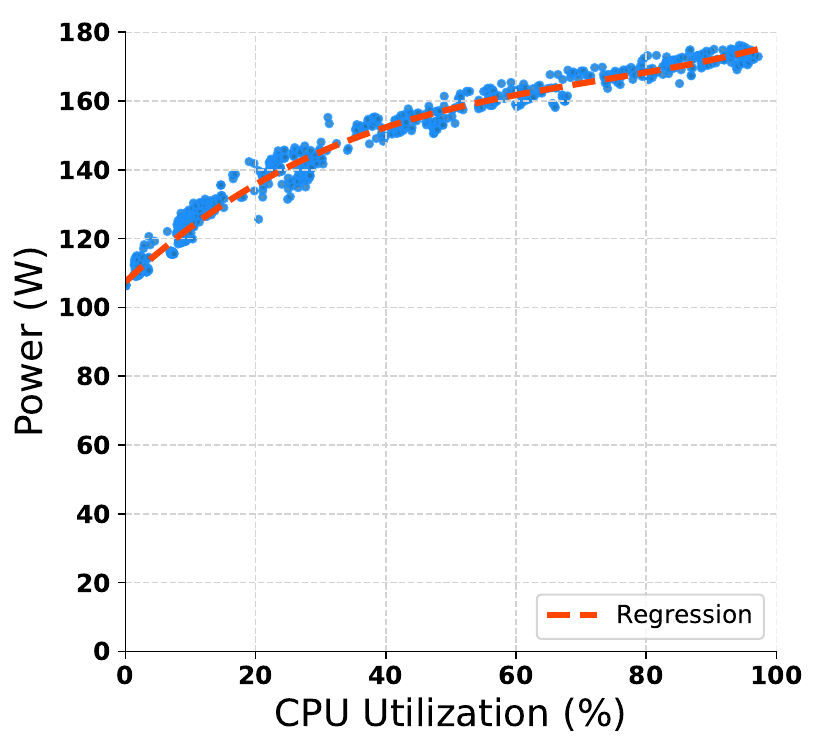}  &
\includegraphics[width=0.5\linewidth]{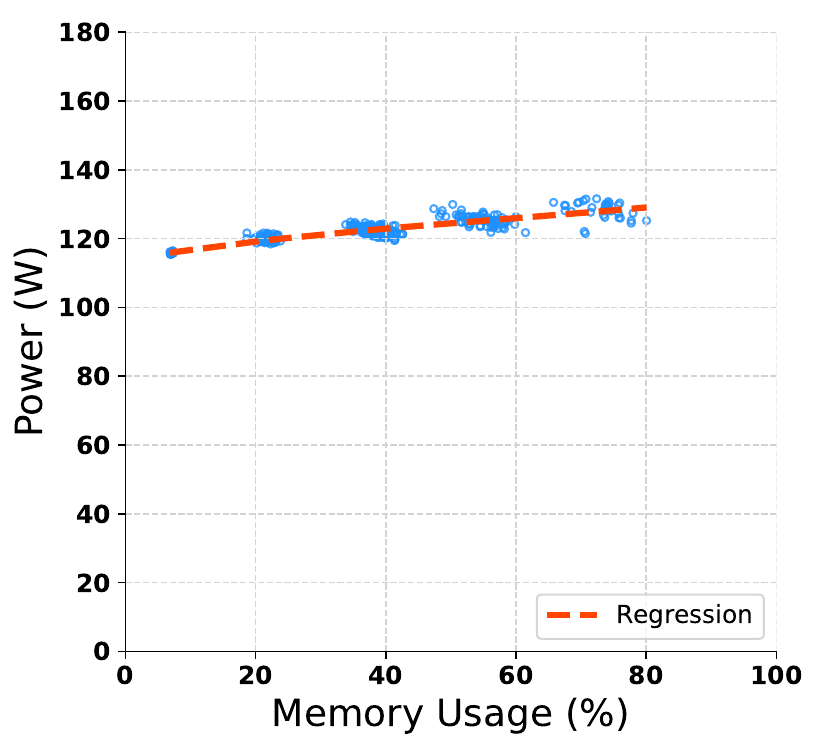} \\
(a) Power vs. CPU Utilization & (b) Power vs. Memory Usage
\end{tabular}
\vspace{-0.2cm}
\caption{\textbf{\emph{Relationship between power consumption of a server and its CPU utilization (at fixed memory usage) and memory usage (at fixed CPU utilization) when replaying production workload traces on a physical server and monitoring power consumption using an external power meter.}}}
\label{fig:example-cpu-power}
\vspace{-0.5cm}
\end{figure}
  
\vspace{0.1cm}
\textbf{Power Model.} A traditional server consists of multiple components that consume power including CPUs, memory, and  I/O devices, e.g., network card, disk, etc.  Prior work shows that a traditional server's power consumption primarily depends on its CPU utilization~\cite{power-api}, since the contribution from other components is nearly constant and not dependent on their usage.  However, as memory sizes in servers increase to support data-intensive applications and memory technology improves to provide a more dynamic power range, memory power consumption is becoming both a significant part of system-level power and also usage-dependent.  As a result, to construct our power usage traces, we use both CPU and memory usage information in the traces above to estimate server power consumption. 

To derive our server power consumption traces, we conduct an empirical study that maps a job's CPU utilization and memory usage to its power consumption on an example physical server.  We randomly sample 10,000 usage readings from each of the Azure and Google traces.  We then replay them on our physical server connected to an external power meter that records server-level power consumption.  To replay traces, we use the stress-ng tool~\cite{stress-ng} that stresses a server's CPUs, memory, and network interface based on the time-varying resource usage information provided in the workload traces. We only stress CPU and memory as both traces only provide CPU and memory usage information and measure the resulting power consumption of the server under that workload.

Figure~\ref{fig:example-cpu-power}(a) shows the actual measurements of power when only the CPU was stressed, along with a cubic polynomial fit to the data using the ordinary least squares method.  We note that even though power and CPU utilization exhibit a clear relationship, it is non-linear and will vary across different types of servers. Figure~\ref{fig:example-cpu-power}(b) then shows the actual measurement of power at varying memory usage at a fixed CPU utilization. We can see that the power consumption varies both with CPU and memory usage.  We sample power data from the results of these experiments to define a power model, which converts usage information in our traces to power consumption. For example, if a job has a 50\% CPU utilization and consumes 1GB of memory, we sample a random data point from the previous experiments that were run with these configurations. The variations in power for a given configuration are due to the use of other server components. 

{\color{black} Figure~\ref{fig:example-cpu-power} also illustrates that, in general, servers are not energy-proportional, and thus may consume substantial, roughly static baseload power when idle.  In Figure~\ref{fig:example-cpu-power}, the baseload power (105W) is $\sim$60\% of peak power (175W).  The baseload-to-peak power ratio varies widely across servers, generally between 30-70\%. Our work focuses primarily on attributing a server’s marginal power, i.e., between its baseload and peak power, to applications based on their resource usage, since attributing baseload power is largely a subjective policy choice. Our dissagregation approach is compatible with any policy. As we discuss in \Section\ref{sec:design}, we attribute a server’s baseload power to applications in proportion to their resource usage (at any give time). However, another policy choice might be to first remove baseload power, and attribute it equally to all applications (regardless of their resource usage).}

\subsection{Qualitative Analysis}

Using the power consumption traces we construct above, we analyze the workload characteristics relevant to disaggregation accuracy including power usage \emph{variability}, \emph{regularity},  and \emph{intensity}. 

\vspace{0.1cm}
\textbf{Variability} refers to the extent or degree of fluctuation or variation in the power consumption of a given job over time. Our evaluation results in \S\ref{sec:evaluation} show that variability is one of the most important factors in determining disaggregation accuracy. This is intuitive: if a job has a non-variable, or constant, power consumption pattern, it is simple to disaggregate, as a model need only learn this constant pattern. We quantify variability using the Coefficient of Variation (CoV), which is defined as the ratio between the standard deviation of a time series over its mean. CoV can have values between 0 and $\infty$ with a CoV greater than 1 typically considered high, i.e., with a standard deviation greater than the average.  

\begin{figure*}[t]
\centering
\begin{tabular}{@{}c@{}c@{}c@{}@{}c}
\includegraphics[width=0.5\linewidth]{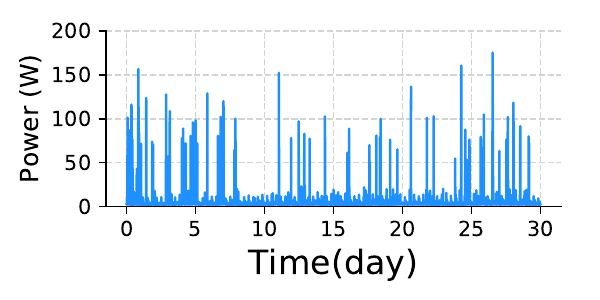} &
\includegraphics[width=0.5\linewidth]{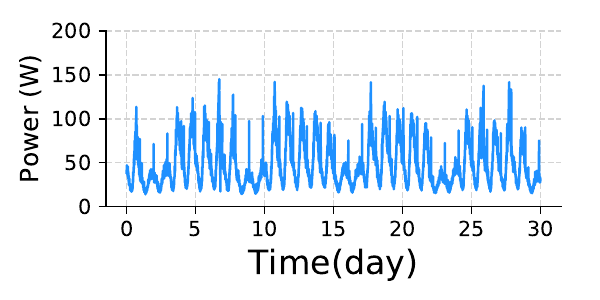} & \vspace{-0.2cm} \\
(a) CoV=$2$  & (b) CoV=$0.5$ \vspace{-0.1cm}  \\
\includegraphics[width=0.5\linewidth]{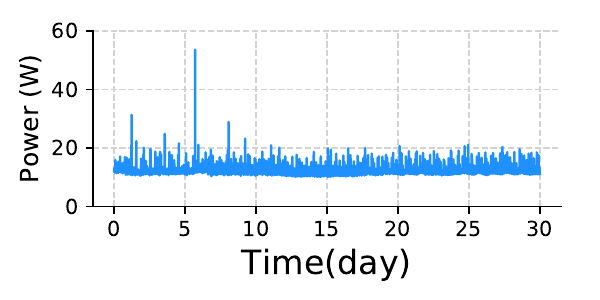} &
\includegraphics[width=0.5\linewidth]{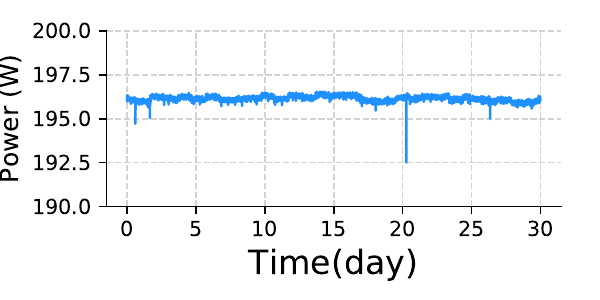} \vspace{-0.2cm} \\
 (c) CoV=$0.1$ & (d) CoV=$0$
 \vspace{-0.2cm}
\end{tabular}
\caption{\emph{\textbf{Illustrative time-series of power consumption for jobs (a-c) with different coefficient of variation (2, 0.5, 0.1), along with an example of a job with $0$ coefficient of variation.} 
}}
\label{fig:example-cvs}
\vspace{-0.5cm}
\end{figure*}

To illustrate, Figure~\ref{fig:example-cvs} shows example time-series of average power (on the $y$-axis) for four jobs in the Azure trace with different values of coefficient of variation of $2$, $0.5$, $0.1$, and $0$.  As expected, high CoV values translate to more frequent and larger variations in power usage, although, as the figure shows, these variations are not necessarily random. 
For a CoV of $2$ (a), the power usage is highly random and does not repeat with any specific pattern. In contrast, a job with CoV of $0.5$  (b) exhibits a distinct pattern of variability in power usage and appears to have a regular pattern that repeats over both 24 hour and 7 day intervals. 
Of course, a lower CoV does not always indicate a regular pattern of usage.  Indeed, the job with CoV of $0.1$ (c) does not exhibit any repeated pattern of usage, and is more volatile than the job with CoV of $0.5$  Finally, a CoV of $0$ (d) indicates a constant power consumption, such that jobs with low CoV values can be more easily disaggregated with higher accuracy. 

CoV is only one metric that correlates with disaggregation accuracy.  We next look at the regularity of the power consumption, which is also related to disaggregation accuracy. 

\vspace{0.1cm}
\textbf{Regularity} refers to the degree to which a given job's power consumption follows a repeating \emph{pattern} over time. Prior work on building energy disaggregation shows that a variable time-series with periodic behavior improves disaggregation accuracy, regardless of its variability. If the pattern of power consumption is perfectly regular and always repeats the same pattern at regular intervals, e.g., every 24 hours, then a model need only learn this pattern to disaggregate a job's power consumption. 

To quantify regularity, we use time-series decomposition that distills our power usage time-series data into its trend, seasonality, and residual (or noise) components, and then apply period detection to the seasonal component~\cite{period-detection}.  The seasonal component represents patterns in the data that repeat over time, while the time between these repeated patterns represents the period. Of course, our time-series data is noisy such that similar, but not exact, patterns of power usage may repeat, and at periodic intervals that vary slightly. Thus, given the noise in the power usage data, simply translating it into the frequency domain and applying a threshold or using autocorrelation is not sufficient for accurate period detection, as discussed in prior work~\cite{predictions}. Specifically, application power usage, even when periodic, is often noisy with many interruptions and random load periods; in addition, periodic behavior also often exhibits increasing or decreasing trends with potentially wide variations in the peaks and troughs power usage~\cite{predictions}.  

\begin{figure*}[t]
\centering
\begin{tabular}{@{}c@{}c@{}c@{}@{}c}
\includegraphics[width=0.5\linewidth]{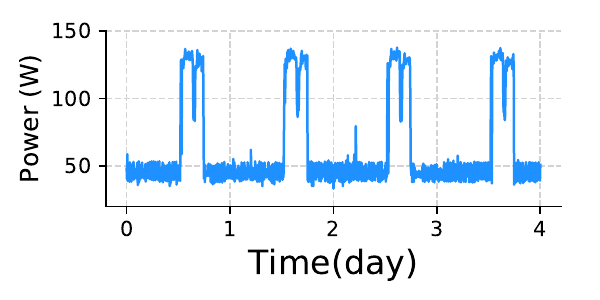} &
\includegraphics[width=0.5\linewidth]{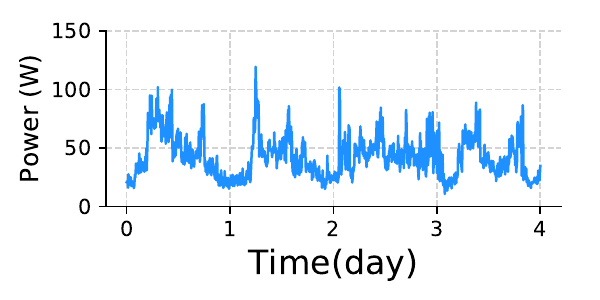} \vspace{-0.2cm} \\
(a) $24$-hour, $0.9$  & (b) $24$-hour, $0.5$ \vspace{-0.1cm}
\\
\includegraphics[width=0.5\linewidth]{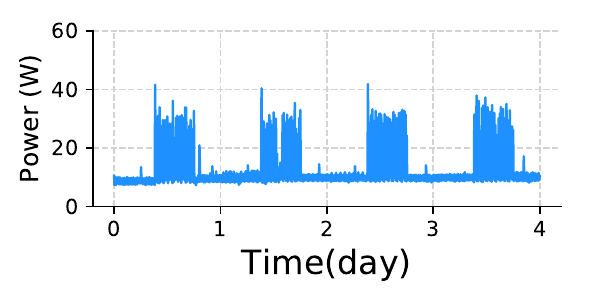} &
\includegraphics[width=0.5\linewidth]{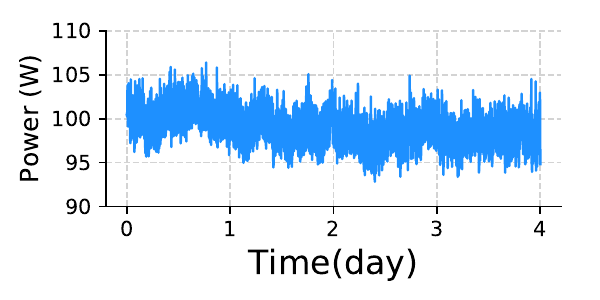} \vspace{-0.2cm} \\
(c) $24$-hour, $0.1$ & (d) No period, $0$  \vspace{-0.1cm} \\
\end{tabular}
\caption{\emph{\textbf{Illustrative time-series of power consumption for jobs with detected periods of length 24-hours (a-c) with different scores (0.9, 0.5, 0.1), along with an example of a job with no period and a score of $0$.}}}
\label{fig:example-scores}
\vspace{-0.3cm}
\end{figure*}

\begin{figure*}[t]
\centering
\begin{tabular}{@{}c@{}c@{}c@{}@{}c}
\includegraphics[width=0.335\linewidth]{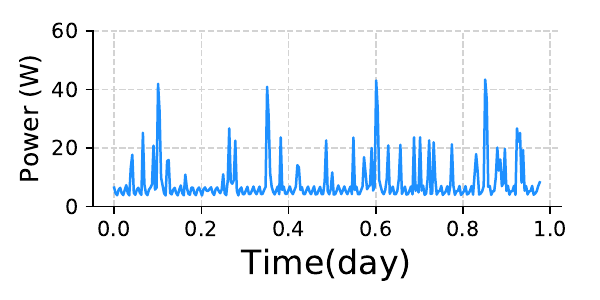} &
\includegraphics[width=0.335\linewidth]{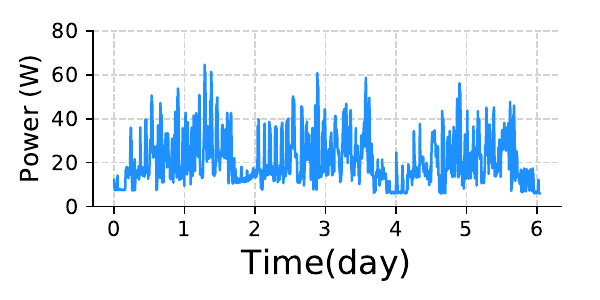} &
\includegraphics[width=0.335\linewidth]{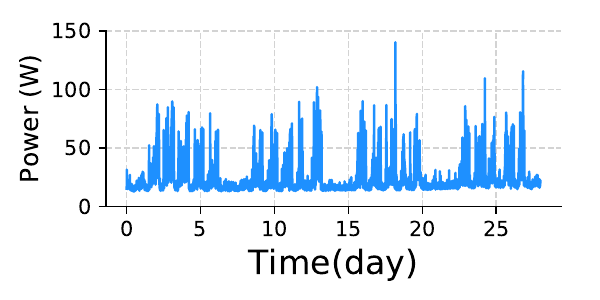} \\
\vspace{-0.2cm}
(a) $6$-hour, $0.5$  & (b) $2$-day, $0.5$  & (c) $7$-day, $0.5$ 
\end{tabular}
\caption{\emph{\textbf{Illustrative time-series of power consumption for jobs with detected periods of length $6$-hours, $3$-days, and $7$-days, all with scores of $0.5$.}}}
\label{fig:example-periods}
\vspace{-0.5cm}
\end{figure*}

Given the size of the dataset, we leverage builtin functions in the Azure Data Explorer~\cite{azure-data-explorer} tool to detect regular periods in our power consumption, which includes an optimized implementation of the basic time-series decomposition functions above.  Specifically, we used the {\tt series\_periods\_detect()} function in Azure DataExplorer~\cite{periods-detect}.   This period detection algorithm detects time-series periods and assigns them a score in the range $[0,1)$ where higher values indicate more intense and regular periodicity, i.e., with less deviation and noise in both the pattern and interval of repetition.  The algorithm reports any periods detected with a non-zero score, and in most cases, it reports many periods for any given time-series.  For example, a time-series that has a 4-hour period likely also has a 24-hour period, although variance in the pattern and interval of repetition may cause the 24-hour period to have a different score. In general, we focus our analysis on the most dominant period that has the highest score.  Finally, our time-series decomposition analysis also assigns jobs without any periods a score of $0$. 

To illustrate, Figure~\ref{fig:example-scores} shows power usage time-series for four jobs that have the same 24-hour period, but with different scores of $0.9$, $0.5$, $0.1$, and $0$.  The figure shows how high scores translate to both high similarity in the pattern of power usage along with the interval between the patterns.  Note that we call the repetitive pattern of power usage the \emph{power signature}. For a score of $0.9$ (a), the power signature is relatively simple, and nearly the same each time, and repeats at nearly precise 24 hour intervals.  In contrast, a job with a $0.5$ score (b) is more variable: there is clearly a repetitive pattern roughly every 24 hours (and also every 4 hours and at even smaller intervals), but the magnitude of the power signature, while often similar, exhibits some distinct variability.  In particular, there is a large spike on the second day along with some smaller variations across the other days.  Similarly, the job with $0.1$ score (c) exhibits even more noise with a less apparent 24-hour period, while the job with $0$ score (d) has no discernible period and appears to be random noise.  Similarly, Figure~\ref{fig:example-periods} shows data from jobs with the same score of $0.5$, but for different periods. This figure demonstrates that time-series decomposition can recognize a range of different period intervals. 

The figures above show that high periodicity scores translate to power signatures that repeat at a regular periodic interval, such that a higher score represents more similarity and regularity with less noise.  As the score decreases, the similarity in the power signatures and strength of the periodicity both decrease, but are still clearly evident, while the noise level, i.e., variability, increases.  These empirical observations of periodicity on these and other jobs in our dataset indicate that any positive score represents some periodicity in the signal that is potentially useful in disaggregating application power usage.  Likewise, Figure~\ref{fig:example-scores}(d) shows that a $0$ score indicates a random or noisy power with no discernible regular pattern of usage.

\begin{figure*}[t]
\centering
\begin{tabular}{@{}c@{}c@{}c@{}@{}c}
\includegraphics[width=0.335\linewidth]{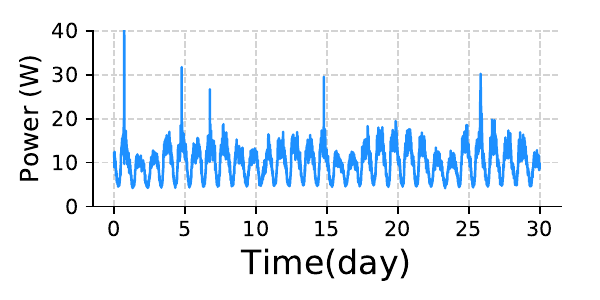} &
\includegraphics[width=0.335\linewidth]{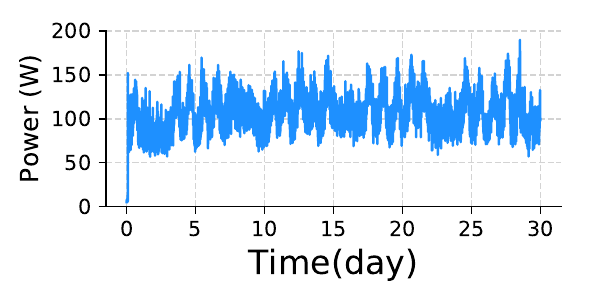} &
\includegraphics[width=0.335\linewidth]{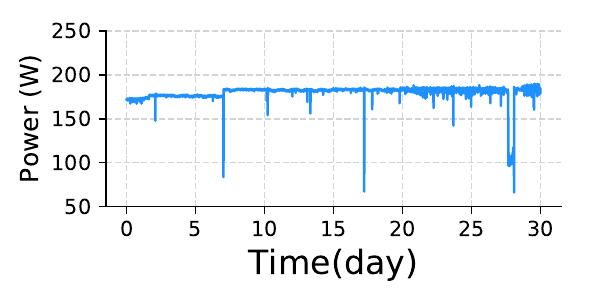}\vspace{-0.1cm}  \\
\vspace{-0.2cm}
(a) $10$ watts  & (b) $100$ watts  & (c) $180$ watts 
\end{tabular}
\caption{\emph{\textbf{Illustrative time-series of power consumption for jobs with average magnitude of $10$W, $100$W, and $180$W.}}}
\label{fig:example-intensity}
\vspace{-0.5cm}
\end{figure*}

\vspace{0.1cm}
\textbf{Intensity} refers to a job's average magnitude of power consumption. We quantify intensity using a job's average power with a range between 0 and the maximum server power. A high or low value of average power is better for disaggregation. 
A job with very high or very low intensity is easier to disaggregate compared with a medium-intensity job, since the high/low-intensity jobs have less room for variation in their power consumption.  That is, if the average power consumption is near the server's maximum or minimum power it means that any deviations from the average must be brief. 

To illustrate, Figure~\ref{fig:example-intensity} shows example time-series of power usage (on the $y$-axis) for three jobs that have different average power magnitudes of $15$W, $60$W, and $180$W.  The figure shows that both high and low magnitudes (a,c) have less room for variation and a less dynamic range of power consumption. As a result, the power consumption patterns at both extremes are, almost by definition, relatively constant.  However, the average magnitude values, e.g., $100$W (b), can come from highly variable power usage patterns, which makes accurate disaggregation more challenging.

\subsection{Large-scale Quantitative Analysis}

In this section, so far, we have defined the various characteristics of real-world workloads that can impact disaggregation accuracy and presented illustrative figures to develop an intuitive understanding of each metric. Below, we present results quantifying the presence of these characteristics in real workloads. 

\begin{figure*}[t]
\centering
\begin{tabular}{cc}
\includegraphics[width=0.4\textwidth]{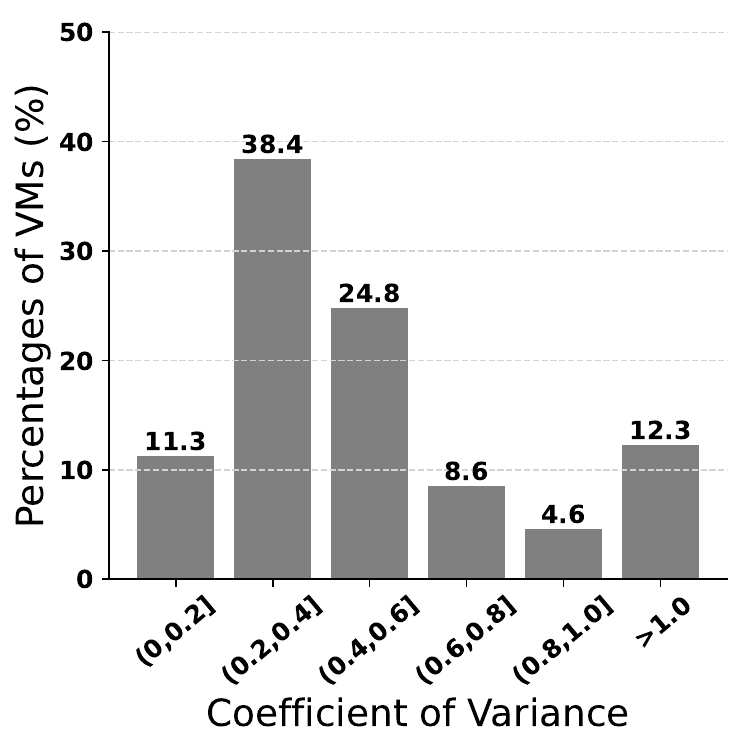} &
\includegraphics[width=0.4\textwidth]{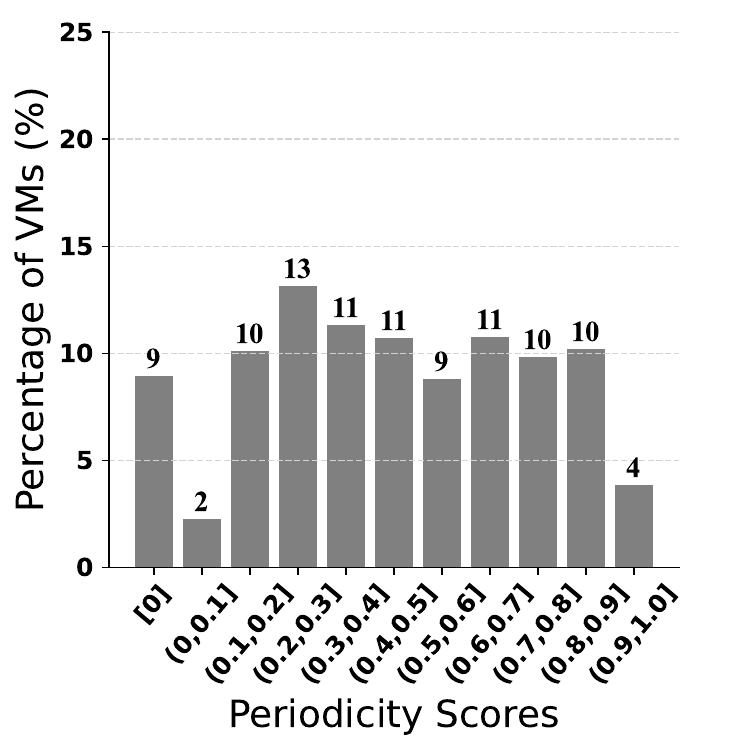} \\
(a) Coefficient of Variation Distribution & (b) Periodicity Score Distribution \\
\includegraphics[width=0.4\linewidth]{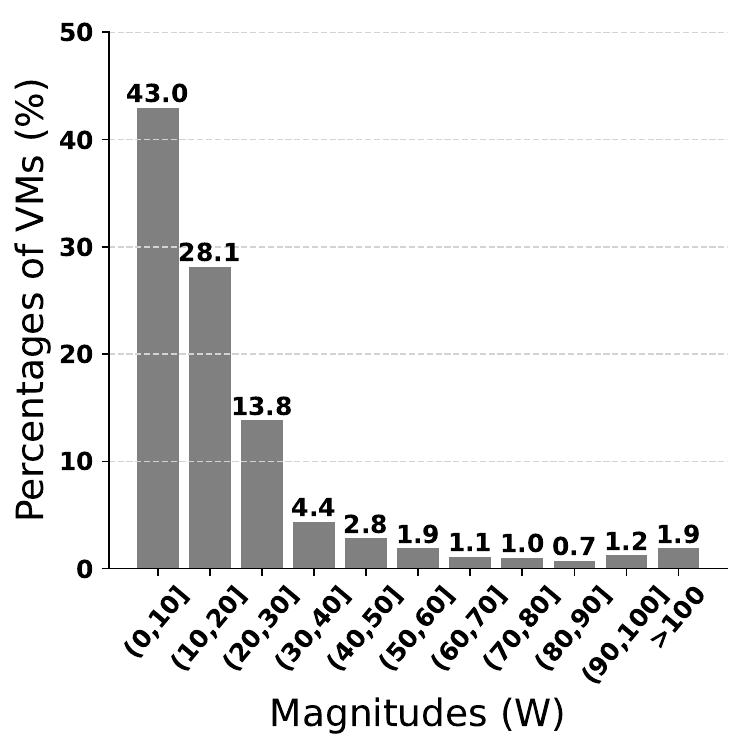} &
\includegraphics[width=0.4\linewidth]{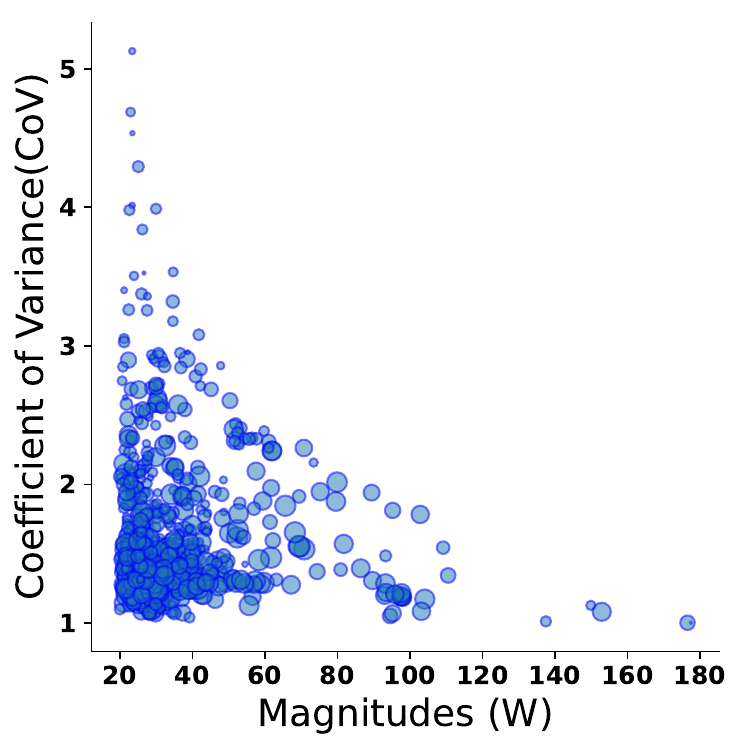} \vspace{-0.1cm} \\
 (c) Magnitude Distribution & (d) Combined
\end{tabular}
\vspace{-0.1cm}
\caption{\emph{\textbf{Distribution of coefficient of variance (a), periodicity score (b), and magnitude (c) from a random sample of 10,000 jobs in Azure workload trace.  The analysis shows that job power consumption at large scales is less variable and highly periodic.  The average power consumption is also low and consistent across jobs regardless of their CoV or score. Finally, (d) shows the CoV, periodicity score, and magnitude for for 1000 jobs to illustrate most jobs score low on CoV and magnitude.}}}
\label{fig:distributions}
\vspace{-0.5cm}
\end{figure*}

Figure~\ref{fig:distributions}(a) shows a histogram of the CoV for each quintile between 0 and 1 for 10,000 jobs randomly sampled from the Azure trace, as well as the percentage of jobs with CoV greater than 1.  In general, CoV's below 1 are considered low, i.e., with standard deviation less than the mean, and those above 1 are considered high.  The graph shows that the vast majority (74.5\%) of jobs in the Azure trace have CoV's less than 0.6.  While most of these jobs have some variation (63.2\% have CoVs between 0.2 and 0.6), it is generally low.  In addition, only 12.3\% of jobs have high CoV's greater than 1 that would make accurate disaggregation especially challenging.  Thus, our large-scale analysis of variability indicates that the vast majority of jobs have low variability that is amenable to accurate disaggregation. Figure~\ref{fig:distributions}(b) next shows a histogram of the regularity in job power usage, where the $x$-axis represents periodicity scores in deciles, and the $y$-axis is the fraction of jobs with their highest periodicity score in that range. The analysis shows that over 91\% of jobs exhibit some non-zero periodicity with over half of jobs exhibiting strong periodicity scores above $0.5$. In contrast, only 9\% of jobs exhibit no detectable periodic behavior in their power usage.  
Next, Figure~\ref{fig:distributions}(c) shows the distribution of average power consumption for the jobs. The graph shows that most jobs have very lower power consumption; 43\% have less than 10W power consumption, assuming they run on a server with 200W maximum power, and 84.9\% have less than 30W power consumption. These power consumption values correspond to roughly 5\% and 15\% resource utilization on a 200W server. 

Finally, in addition to the individual distributions, we show the distribution of CoV, periodicity, and magnitude for a randomly sampled 1000 VMs from the 10,000 jobs in Figure~\ref{fig:distributions}(d).   In this graph, the magnitude of average power consumption is on the $x$-axis, the CoV is on the $y$-axis, and the size of each datapoint represents the periodicity score.  The overall takeaway is that majority of jobs in our trace have low magnitude and CoV, while also frequently exhibiting regular and periodic patterns of power consumption.  As a result, real-world workloads are highly amenable to  energy disaggregation. In particular, our analysis above indicates that, while server applications do not have to exhibit regularity in their power usage, at production scales they tend to be highly regular with little noise.  This is likely due to the fact that most jobs at large scales are deployed to serve a specific purpose and type of workload with a regular pattern of power usage.  In addition, the periodic intervals cover a wide range, which indicates that different jobs have widely different patterns of power usage.  In addition to distinctive periods, real jobs also tend to have distinctive power signatures during their active periods, i.e., the magnitude and pattern of power usage when active, relative to other jobs. Yet, these distinctive power signatures for each job tend to be highly similar across different active periods.  The example jobs in Figures~\ref{fig:example-scores} and \ref{fig:example-periods} illustrate both of these characteristics, i.e., distinctive power signatures across different jobs but similar within the same jobs.

\subsection{Implications of Analysis}

Our analysis above demonstrates not only is there significant potential to disaggregate application-level server power, but production workload characteristics suggest that disaggregation may actually be \emph{much more effective} when disaggregating server power compared to its original use in disaggregating building power into individual electrical loads for numerous reasons. Specifically, while many large electrical loads exhibit periodicity, they are often thermostatically-driven, so the periodic interval varies based on environmental conditions or user behavior.  In contrast, from our analysis, many jobs' power usage tend be have deterministic periods, i.e., likely in some cases driven by timers as cron jobs.  In addition, job power signatures tend to be highly distinctive and thus identifiable in the aggregate power due to the wide variability in how applications exercise resources. In contrast, electrical loads typically exercise power in highly similar ways, which makes distinguishing them in the aggregate power signal $P(t)$ more challenging.  For example, the large majority of electrical loads consist of either resistive heating elements (e.g., coffee makers, toasters, ovens, etc.), motors (e.g., vacuums, AC compressors, fans, etc.), or both (e.g., electric dryers), which have similar power signatures~\cite{barker:jsac,barker:igcc13}.  Finally, given our analysis above, there will likely be few co-located jobs on any server that either have no periods, or have the same overlapping period interval (which would reduce disaggregation accuracy).  Further, even if multiple ``noisy'' jobs with no period were co-located, their power usage is likely to be low, and thus likely to only minimally affect the accuracy of disaggregating other jobs that oscillate between periods of high power usage and low power usage.

\section{\systemName Design}
\label{sec:design}
In this section, we present the design  \systemName, our system for non-intrusively monitoring application-level power consumption by disaggregating power data from external server- and rack-level power meters. 
Below, we describe \systemName's overall system architecture and its different components for training disaggregation models and using them to disaggregate external power data. 

\begin{figure}[t]
\centering
\includegraphics[width=0.95\linewidth]{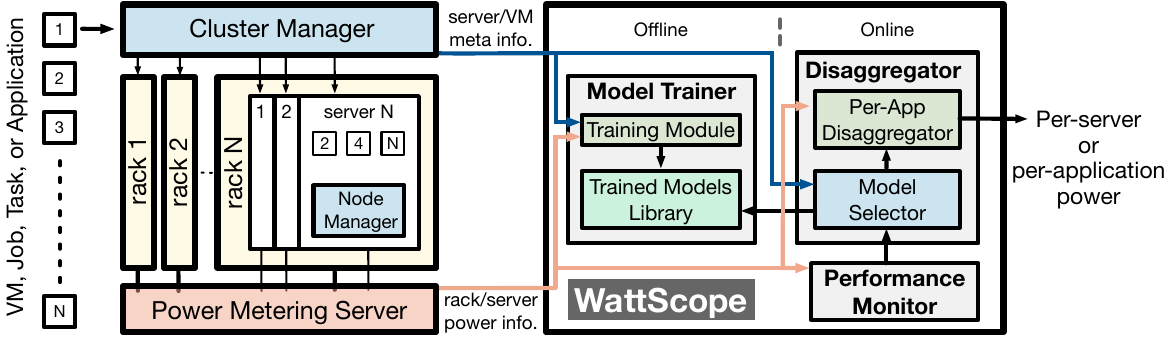} 
\vspace{-0.1cm}
\caption{\textbf{\systemName\emph{overview and its three key components: model trainer, disaggregator, and performance monitor.}}}
\label{fig:wattscope-overview}
\vspace{-0.5cm}
\end{figure}


Figure~\ref{fig:wattscope-overview} shows \systemName's architecture for non-intrusive application-level power monitoring. \systemName is implemented as a cluster-level system for monitoring application-level power in datacenters that does not require any hardware or software support on the servers running the applications.  The only requirement is a network-accessible external power meter that monitors each server's power, which most datacenters already have.  As in a typical datacenter, \systemName assumes users submit their workload to a cluster manager in the form of individual jobs or tasks, which run inside containers or VMs. The cluster manager schedules the jobs on one of the servers depending on resource availability and job placement constraints. The cluster manager, or the node manager, keeps track of the high-level category and placement for each job, such as their scheduling priority, nature of the job (service, batch, or interactive), and user information. All datacenters need such information for scheduling and billing purposes. Optionally, the cluster manager may collect information on the resource usage for all the jobs, such as CPU utilization and memory usage. While such information is {\em not necessary for disaggregation}, \systemName can opportunistically use it during the training process if available. There is also an external power monitoring server that records the power consumption at the server- or rack-level over a pre-defined sampling interval $\Delta$$t$ and exposes that information through an API. \systemName takes the server- or job-level meta information and rack- or server-level aggregate power consumption and reports the server- or job-level power consumption. While \systemName can disaggregate rack-level power consumption into servers, we focus primarily on disaggregating server-level power into application-level power consumption information in the remainder of this section.

Importantly, \systemName assumes that resource allocations for applications are reserved and not best-effort.  If resources are allocated best-effort, then applications' variations in resource usage and power is a function of not only their own behavior, but also the behavior of co-located applications.  In this case, the scheduler would dictate applications' resource and power variations rather than their own behavior, which would prevent training our models below.  However, our assumption of reserved resources generally holds for production schedulers in industry, such as Borg~\cite{borg}. While production schedulers do overcommit resources, i.e., by reserving allocating more resources than exist on a server, they attempt to minimize application throttling, i.e., where applications attempt to use their reserved resources but they are not available. Prior work shows that production schedulers rarely throttle applications~\cite{limit-eurosys}. As a result, a cluster manager's effectiveness at packing jobs onto servers has no impact on the variability of a job’s resource usage.

In addition, the type of server a job runs on also affects its resource usage characteristics and power, which also affects the efficacy of disaggregation models. In many cases, since jobs are not throttled, changing server types will only alter the magnitude of the resource usage and not its variations. In addition, datacenters often operate large clusters of homogeneous servers, which facilitates training models for each server family. In general, there is a tradeoff between model accuracy and modeling cost.

\systemName consists of three key components (or modules): an offline model trainer, an online disaggregator, and an online performance monitor. Below, we describe the function of each module in detail. 

\subsection{Model Trainer}
\label{sec:model-trainer}
The model trainer module's task is to train a library of models that can then be used by the disaggregator. While the model trainer can be modified to work as an online module, we design it as an offline module that trains and stores multiple models. The model trainer takes three inputs: (i) ground truth application-level power usage, (ii) aggregate server-level power usage, and (iii) meta information about the applications. 

\vspace{0.1cm}
\textbf{Inputs.}
First, to train an energy disaggregation model, we need ground truth application-level power data. However, as discussed in \S\ref{sec:workload}, physically monitoring per-application power usage is not possible. To solve this problem, we use alternative methods that provide approximate power consumption with varying levels of accuracy. In particular, our approach is to use data collected by an intrusive software-based method for application-level power monitoring, such as PowerAPI~\cite{power-api}.  This data can be collected in the same datacenter on a subset of machines running the representative workload or in another datacenter that has similar workload characteristics. However, such fine-grained per-application power monitoring is not deployed in practice.  Thus, another option is to use the resource usage information, such as CPU and memory utilization, as a proxy for estimating power consumption using a power model, such as the one we used in \S\ref{sec:analysis-setup}.  Second, we need the aggregate power consumption information for the server, which is typically collected in datacenters for power management, billing, and cooling purposes. Third, the metadata information about the servers and applications is used as a key for distinguishing trained models, which can later be used by the disaggregator to select a model depending on the characteristics of the workload running on the server. This information can include job type, hashed user information, job priority, or any other information that can be used to identify a given job or class of jobs.

\vspace{0.1cm}
\textbf{Training.}
Given our problem's similarity to energy disaggregation in buildings, we examined numerous existing approaches from the domain of building energy disaggregation \cite{nilmtk}.  In general, building energy disaggregation techniques require a per-load model that captures characteristics of each load's pattern of energy usage.  These models were initially simple and pre-configured, e.g., by specifying a small number of discrete power states for each load, based on \emph{a priori} knowledge of each load.  However, recent approaches have instead captured loads using machine learning models, e.g., neural networks, trained on datasets of buildings where each load's power is separately monitored to provide ground truth~\cite{nilmtk,nilmtk-contrib}.  While we evaluate many of these approaches in \Section\ref{sec:implementation}, we adapt and extend a recently proposed sliding window approach that uses deep neural networks (DNNs) as the basis for \systemName~\cite{sliding-window}.  As we show in \Section\ref{sec:implementation}, this technique provides the highest accuracy, in part, because it is best suited for the characteristics of loads like server applications that have multiple (or continuous) power states.

Specifically, our sliding window approach takes a window $w$ of data points as input that represent the past $w$ samples of a server's aggregate power, e.g., $[P(t-100),P(t-99),...,P(t-1),P(t)]$. As discussed in \Section\ref{sec:workload}, since we use the server's aggregate power, rather than its marginal power, our approach implicitly attributes a server’s baseload power to applications in proportion to their resource usage (at any give time). This input feeds into a convolutional layer, which in-turn feeds into two bidirectional GRU (Gated Recurrent Unit) layers and two dense layers, such that dropout units are inserted between these layers.  Each of the layers uses an ReLU (rectified linear unit) activation function except the last dense layer, which uses a linear activation function. Ultimately, the output is the disaggregated power usage $p_i(t)$ for load $i$ at time $t$. Prior work has shown that replacing the GRU layers with LSTM layers results in similar accuracy across a wide range of loads, but requires both more memory and computation for training.  The model inserts dropout units, which probabilistically drops outputs, to both prevent over-fitting and improve robustness with respect to missing values.  The window size is generally set to $50-100$ datapoints, although the optimal window $w$ may vary for different loads. We discuss our model's specific implementation details in \Section\ref{sec:implementation}, including the configuration and hyperparameters of each layer.  Note that a specific application's load model is trained using data from multiple servers, which may operate different sets of applications with different characteristics. 

\vspace{0.1cm}
\textbf{Model library.}
The set of background applications running on a server can significantly vary, over time and across different servers, in terms of the number of applications and their characteristics. As a result, it is not possible to train a model for each combination. If our power usage trace provides information on the co-location of various applications, we train models for the most common co-location scenarios. There is a trade-off between the number of models trained and disaggregation accuracy; the higher the number of models, the higher the accuracy and vice versa.  If the power usage trace does not provide co-location information, such as in our Azure-based power trace, we randomly select different applications on a server and train models for wide range of application combinations.  All of the models are indexed by the metadata information about the applications whose data was used to train a given model. For example, a model trained with applications that have low variability, low regularity, and medium intensity is saved with this information as a label, enabling it to be selected by the disaggregator's model selector for applications with similar characteristics.

\subsection{Disaggregator}

Formally, our disaggregation problem can be stated as follows: given a certain number of applications running on a server, as indicated by the cluster-level scheduler, and the server's aggregate power consumption $P(t)$ at time $t$, 
we need to infer the average power consumption $p_i(t)$ (over a sampling interval $\Delta$$t$) attributable to each application $i$.  \systemName operates in real-time by inferring each application's average power usage over $\Delta$$t$ immediately after the external power meter reports a new power sample for the server.  While we assume a 5-minute sampling interval $\Delta$$t$ to match the resolution of our production trace, our approach is applicable to any sampling interval on the order of seconds-to-minutes. Note that our approach can also disaggregate average power usage (or equivalently energy) over coarser time intervals than the sampling interval by simply averaging the inferred power usage over the coarser interval, e.g., to infer an application's energy usage over a month for billing purposes.  In general, the longer the interval, the more accurate the disaggregation. 

The model selector component of the disaggregator selects a model from the library for disaggregation depending on the characteristics of the applications running on the server. As the metadata information provides high-level information for the applications, collectively used as a label for a model during the training phase, the model selector uses that information to pick an appropriate model. As the current combination of applications on the server may not exactly match any of the trained models, the model selector chooses the closest model to use for disaggregation. The metric used to quantify ``closeness'' is subjective and depends on the operator's choice.

\subsection{Performance Monitor}
The performance monitor module keeps track of the currently deployed disaggregation model and sends feedback to the model selector if the accuracy of the current model starts to degrade. To quantify the performance of a given model, it compares the total allocated power to the applications with the ground truth aggregate power consumption. Under normal circumstances, the error should be under some pre-defined threshold. However, the error will increase if the number of applications or their characteristics change. If a high error persists for more than a specified period of time, it sends a signal to the model selector to select a new model for disaggregation.

\section{Implementation}
\label{sec:implementation}
We implemented \systemName and seeded it with multiple job models trained using data from large-scale Azure and Google production workload traces described in \Section\ref{sec:workload}.  The Google trace includes job co-location information, i.e., which jobs are co-located on the same server, while the Azure trace does not. Since the Azure trace only includes job-level average resource usage statistics, every 5 minutes and not server-level placement information or power data, we use the trace to construct our own synthetic ground truth power data for training. Specifically, we assume all jobs run within VMs (or containers), and each is given an equal number of resources on a server and not throttled. To provide some context, we ascribe a maximum power of 200W to each job based on its resource allotment, such that each VM can independently consume up to 200W when operating at full utilization, as in prior work~\cite{VM-Power}.  We also assume that power consumption is related to utilization based on the function from Figure~\ref{fig:example-cpu-power}.  We then construct training datasets by simulating the mixing of different jobs together on the same server, such that the server's total power is 200W$\times$$n$, where $n$ is the number of jobs.  So, for example, a server in our training data that runs 5 jobs has a peak power of 1,000W (or 5$\times$200W). Our contextual parameterization of 200W is arbitrary, and only relevant to experiments that cite power values.  In most cases, we report normalized results that are not dependent on server power. Note that, since we derive our ground truth power data from a resource-power model, our experimental results using this data do not incorporate inaccuracies due to this model, but only quantify inaccuracy due to disaggregation.  There is substantial prior work on accurately modeling the relationship between resource usage and power~\cite{power-api}.  While we leverage this work, our approach is orthogonal to it.

Our approach above is general, and can apply to servers at any level of power consumption.  Also, note that the approach above would not mimic reality if the jobs running on a server consumed the entire CPU, since at that point, they would conflict with each other in a way that would affect their utilization and power consumption.  However, prior work has shown that such conflicts are exceedingly rare, and cluster schedulers include sophisticated algorithms to avoid them even when overcommitting resources~\cite{limit-eurosys}.  

For the Azure trace, we use the approach above to generate a large number of synthetic training datasets for different specific jobs and job classes running on servers with a range of different other jobs and job classes.  As discussed in \Section\ref{sec:design}, the training data set for a particular job takes a prior window $w$ of aggregate server-level power as input, and produces a job's disaggregated power.    In \Section\ref{sec:evaluation}, we evaluate disaggregation accuracy with models trained in a variety of different ways, from more-to-less specific.  We implement a number of disaggregation models by modifying nilmtk-contrib~\cite{nilmtk-contrib}, an open-source toolkit implementation of numerous algorithms for energy disaggregation of buildings. In particular, we replace nilmtk-contrib's existing training data sets with the synthetic training data above, and also eliminate pre-existing configuration meta-data that is specific to particular building loads, e.g., refrigerators, ACs, etc. to make our implementation generic.  The toolkit includes numerous other benchmark algorithms, which we evaluate below. 

We use two primary metrics for evaluating \systemName's accuracy: Mean Absolute Error (MAE) and Normalized Mean Absolute Error (NMAE), shown below.  MAE is simply the average of the absolute difference between the inferred power $p_i(t)$ of VM $i$ and its actual power $\hat{p_i(t)}$ over all times $t$, while the NMAE is the MAE normalized by the job's mean power. 

\vspace{-0.3cm}
\begin{equation}
MAE_i =  \sum_{t=1}^{T} \frac{|p_i(t)-\hat{p_i(t)}|}{T}
\end{equation}

\vspace{-0.3cm}
\begin{equation}
NMAE_i =  \frac{MAE_i}{\frac{1}{T} \sum_{t=1}^{T} \hat{p_i(t)}}
\end{equation}

The MAE is in units of watts (W) and shows the absolute error in \systemName's inferred power, while the NMAE quantifies the error as a percentage of a job's mean power.  In general, low power jobs tend to have higher NMAEs even when their MAE is low in an absolute sense, especially since low power jobs are more challenging to disaggregate from server power that may be much higher.  Likewise, high power jobs may have low NMAEs even if when their MAE may be comparatively high.  Thus, in our evaluation, we contextualize these results relative to standard benchmark approaches.  In particular, we compare with the NMAE and MAE for a baseline approach that infers a job's power is always equal to its mean power over the time interval.  We call this the `Mean' model.  We implemented our \systemName, and trained and evaluated our models, on an Intel Xeon Silver 4214R CPU with 12 Cores at 2.4 GHz and 128 GB RAM.

 \begin{table}[]
\setlength{\tabcolsep}{1.4 mm}{
\begin{tabular}{||c|llllll|llllll||}
\hline\hline

& \multicolumn{6}{c|}{\textbf{MAE (W)}}                    
& \multicolumn{6}{c||}{\textbf{NMAE (\%)}}          
\\ \cline{2-13} 
\multirow{-2}{*}{\textbf{Models}}                                       & \multicolumn{1}{l|}{\textbf{job1}}                           & \multicolumn{1}{l|}{\textbf{job2}}                          & \multicolumn{1}{l|}{\textbf{job3}}                           & \multicolumn{1}{l|}{\textbf{job4}}                           & \multicolumn{1}{l|}{\textbf{job5}}                          & \multicolumn{1}{c|}{\begin{tabular}[c]{@{}c@{}}Aver\\ -aged\end{tabular}} & \multicolumn{1}{l|}{\textbf{job1}}                           & \multicolumn{1}{l|}{\textbf{job2}}                           & \multicolumn{1}{l|}{\textbf{job3}}                           & \multicolumn{1}{l|}{\textbf{job4}}                           & \multicolumn{1}{l|}{\textbf{job5}}                           & \multicolumn{1}{c||}{\begin{tabular}[c]{@{}c@{}}Aver\\ -aged\end{tabular}} \\ \hline\hline\hline
Mean                                                                    & \multicolumn{1}{l|}{20.88}                                  & \multicolumn{1}{l|}{4.95}                                  & \multicolumn{1}{l|}{20.33}                                  & \multicolumn{1}{l|}{41.80}                                  & \multicolumn{1}{l|}{2.78}                                  & 18.15                                                                     & \multicolumn{1}{l|}{36.93}                                  & \multicolumn{1}{l|}{27.46}                                  & \multicolumn{1}{l|}{38.99}                                  & \multicolumn{1}{l|}{29.55}                                  & \multicolumn{1}{l|}{29.71}                                  & 32.53                                                                    \\ \hline
CO                                                                      & \multicolumn{1}{l|}{56.44}                                  & \multicolumn{1}{l|}{23.00}                                 & \multicolumn{1}{l|}{53.85}                                  & \multicolumn{1}{l|}{134.27}                                 & \multicolumn{1}{l|}{9.33}                                  & 55.38                                                                     & \multicolumn{1}{l|}{99.81}                                  & \multicolumn{1}{l|}{127.63}                                 & \multicolumn{1}{l|}{103.28}                                 & \multicolumn{1}{l|}{94.91}                                  & \multicolumn{1}{l|}{99.70}                                  & 105.07                                                                   \\ \hline
\begin{tabular}[c]{@{}c@{}}Exact\\ -FHMM\end{tabular}                   & \multicolumn{1}{l|}{16.40}                                  & \multicolumn{1}{l|}{5.75}                                  & \multicolumn{1}{l|}{18.63}                                  & \multicolumn{1}{l|}{41.28}                                  & \multicolumn{1}{l|}{3.36}                                  & 17.08                                                                     & \multicolumn{1}{l|}{29.00}                                  & \multicolumn{1}{l|}{31.90}                                  & \multicolumn{1}{l|}{35.73}                                  & \multicolumn{1}{l|}{29.18}                                  & \multicolumn{1}{l|}{35.96}                                  & 32.35                                                                    \\ \hline
DAE                                                                     & \multicolumn{1}{l|}{16.79}                                  & \multicolumn{1}{l|}{5.00}                                  & \multicolumn{1}{l|}{18.37}                                  & \multicolumn{1}{l|}{39.54}                                  & \multicolumn{1}{l|}{2.73}                                  & 16.49                                                                     & \multicolumn{1}{l|}{29.70}                                  & \multicolumn{1}{l|}{27.76}                                  & \multicolumn{1}{l|}{35.23}                                  & \multicolumn{1}{l|}{27.95}                                  & \multicolumn{1}{l|}{29.22}                                  & 29.97                                                                    \\ \hline
RNN                                                                     & \multicolumn{1}{l|}{17.85}                                  & \multicolumn{1}{l|}{4.89}                                  & \multicolumn{1}{l|}{19.72}                                  & \multicolumn{1}{l|}{39.43}                                  & \multicolumn{1}{l|}{2.74}                                  & 16.93                                                                     & \multicolumn{1}{l|}{31.57}                                  & \multicolumn{1}{l|}{27.11}                                  & \multicolumn{1}{l|}{37.83}                                  & \multicolumn{1}{l|}{27.87}                                  & \multicolumn{1}{l|}{29.33}                                  & 30.74                                                                    \\ \hline
Seq2Seq                                                                 & \multicolumn{1}{l|}{18.30}                                  & \multicolumn{1}{l|}{4.51}                                  & \multicolumn{1}{l|}{18.09}                                  & \multicolumn{1}{l|}{37.66}                                  & \multicolumn{1}{l|}{2.63}                                  & 16.24                                                                     & \multicolumn{1}{l|}{32.37}                                  & \multicolumn{1}{l|}{25.03}                                  & \multicolumn{1}{l|}{34.70}                                  & \multicolumn{1}{l|}{26.62}                                  & \multicolumn{1}{l|}{28.12}                                  & 29.37                                                                    \\ \hline
Seq2Point                                                               & \multicolumn{1}{l|}{16.31}                                  & \multicolumn{1}{l|}{4.39}                                  & \multicolumn{1}{l|}{17.29}                                  & \multicolumn{1}{l|}{35.08}                                  & \multicolumn{1}{l|}{2.58}                                  & 15.13                                                                     & \multicolumn{1}{l|}{28.84}                                  & \multicolumn{1}{l|}{24.35}                                  & \multicolumn{1}{l|}{33.16}                                  & \multicolumn{1}{l|}{24.80}                                  & \multicolumn{1}{l|}{27.55}                                  & 27.74                                                                    \\ \hline
\rowcolor[HTML]{C0C0C0} 
\textit{\textbf{\begin{tabular}[c]{@{}c@{}}\systemName\end{tabular}}} & \multicolumn{1}{l|}{\cellcolor[HTML]{C0C0C0}\textbf{11.02}} & \multicolumn{1}{l|}{\cellcolor[HTML]{C0C0C0}\textbf{3.76}} & \multicolumn{1}{l|}{\cellcolor[HTML]{C0C0C0}\textbf{13.10}} & \multicolumn{1}{l|}{\cellcolor[HTML]{C0C0C0}\textbf{29.61}} & \multicolumn{1}{l|}{\cellcolor[HTML]{C0C0C0}\textbf{2.39}} & \textbf{11.98}                                                            & \multicolumn{1}{l|}{\cellcolor[HTML]{C0C0C0}\textbf{19.49}} & \multicolumn{1}{l|}{\cellcolor[HTML]{C0C0C0}\textbf{20.87}} & \multicolumn{1}{l|}{\cellcolor[HTML]{C0C0C0}\textbf{25.12}} & \multicolumn{1}{l|}{\cellcolor[HTML]{C0C0C0}\textbf{20.93}} & \multicolumn{1}{l|}{\cellcolor[HTML]{C0C0C0}\textbf{25.56}} & \textbf{22.40}                   \\ \hline\hline

\end{tabular}}
\vspace{-0.2cm}
\caption{\textbf{\emph{Errors in disaggregating the power of five different representative jobs running on the same physical server.}}}
\vspace{-0.5cm}
\label{table:comparison}
\end{table}

\systemName's model trainer leverages a neural network that takes as input a sliding window of aggregate power values to infer a job's power, as discussed in \S\ref{sec:model-trainer}.  We train \systemName's neural network for 50 epochs with a batch size of 1024.  We also optimized the training by fine-tuning the hyperparameters based on prior work~\cite{sliding-window}.  Specifically, our sliding window model uses a window size $w$ of the 100 previous datapoints, i.e., aggregate power values that first feed into a convolutional layer with 16 filters of size 4 with stride and a rectified linear unit (ReLu) activation function; this layer feeds into a bidirectional gated-recurrent unit (GRU) with size 64 and a concat merge mode followed by a drop-out unit with weight 0.5; this layer then feeds into another similar layer from before but of size 128; this layer finally feeds into two dense layers of size 128 (with ReLU activation function) and 1 (with linear activation function), respectively, with another dropout unit of weight 0.5 between them. 

We compared \systemName's approach above with a wide range of different disaggregation models implemented by nilmtk-contrib~\cite{nilmtk-contrib}. Table~\ref{table:comparison} shows that \systemName's approach generally has the highest or near the highest MAE, and is also consistent across five different types of job types. For this experiment, all five of these jobs ran on the same physical server, and we trained each model over 7 days and then tested its accuracy over the remaining length of our trace.   By contrast, the other models have more variable accuracy.  For example, Combinatorial Optimzation (CO) has poor accuracy on job 4, but much better relative accuracy on job 5.  Table~\ref{table:comparison} similarly shows the normalized MAE for the same experiment.  In all cases, \systemName yields the lowest MAE and NMAE when inferring each jobs' disaggregated power.  The table also shows the MAE and NMAE between the actual aggregate power and the inferred aggregate power computed based on the sum of the inferred power of each job.

\section{Evaluation}
\label{sec:evaluation}

\begin{figure*}[t]
\centering
\begin{tabular}{@{}c@{}c@{}c@{}c}
\includegraphics[width=0.51\linewidth]{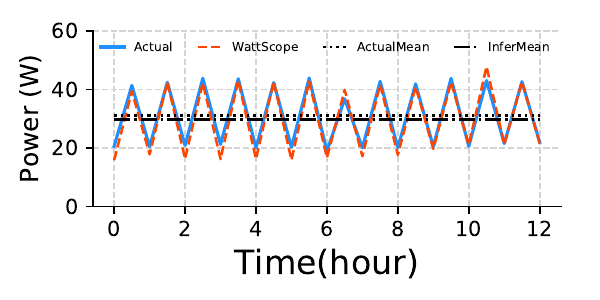} &
\includegraphics[width=0.51\linewidth]{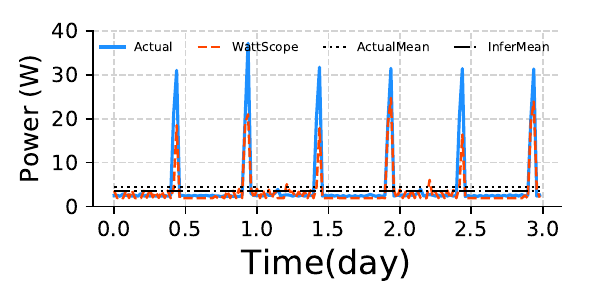}  \vspace{-0.2cm}  \\
\textbf{(a)} $1$h (3.9W, $12$\%)   & \textbf{(b)} $12$h (1.7W, $38$\%) \vspace{-0.1cm} \\
\includegraphics[width=0.51\linewidth]{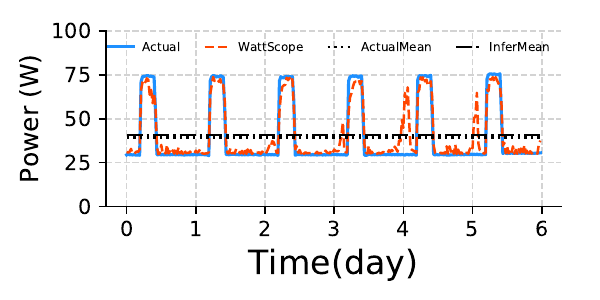} &
\includegraphics[width=0.51\linewidth]{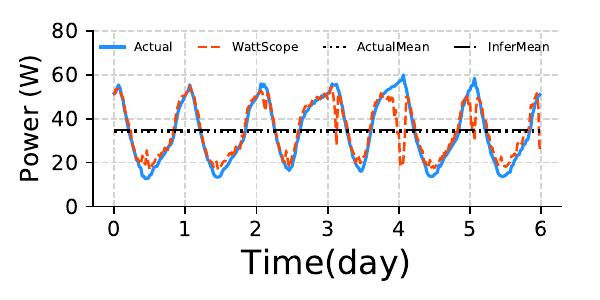}  \vspace{-0.1cm} \\
\vspace{-0.2cm}
\textbf{(c)} $24$h (4.0W, $10$\%)  & \textbf{(d)} $168$h (4.5W, $13$\%) 
\end{tabular}
\caption{\textbf{\emph{Time-series of actual and inferred (disaggregated) power usage of a job for the four representative jobs with different detected periods with high scores, similar coefficient of variations, and different intensities. The caption states the period, mean absolute error ($MAE$), and normalized mean absolute error ($NMAE$).}}}
\label{fig:example-vms}
\vspace{-0.5cm}
\end{figure*}

In this section, we evaluate \systemName for its accuracy in non-intrusively disaggregating total server power consumption into job-level power consumption. We first present qualitative results illustrating the high disaggregation accuracy of \systemName and provide some intuition for our quantitative metrics (\S\ref{sec:qual-results}). 
We next evaluate how job characteristics such as variability, regularity, and intensity affect disaggregation accuracy (\S\ref{sec:job-charac}). We then present quantitative results for desegregating server-level power to job-level power based on actual co-location information in a production trace, which demonstrates how \systemName would work in practice (\S\ref{sec:production}). Finally, we evaluate the \systemName's disaggregation approach for its scalability, robustness, and generalization (\S\ref{sec:scale-robust-general}). Note that for most experiments we use application-specific disaggregation models, i.e., trained on the application's power data.  We quantify the inaccuracy due to using a general model that is not application-specific in \Section\ref{sec:scale-robust-general}.
In evaluating \systemName, we assume that our system knows the characteristics of jobs on a server and uses them to select the appropriate model for disaggregation. Evaluating the performance of our model selector or performance monitor is outside the scope of this paper. 

\subsection{Qualitative Results}
\label{sec:qual-results}
To provide an intuitive meaning to the quantitative results in the following sections, we present the time series of four representative jobs from the Azure trace. Figure~\ref{fig:example-vms} shows the time series of the actual and inferred (or disaggregated) power for each of the four jobs, along with the actual and inferred average power. 
The graphs illustrate that \systemName's disaggregated power is highly accurate and the actual and inferred power closely matches for all of the jobs, as does the actual and mean power. While we choose a more intuitive metric of NMAE for the rest of this section, a high value of NMAE does not necessarily mean poor disaggregation performance. NMAE can be quite high for jobs with low intensity as the 12h job shown in Figure~\ref{fig:example-vms}b has an NMAE value of 38\% due to its average power of less than 5W.

\begin{figure*}[t]
\centering
\begin{tabular}{ccc}
\includegraphics[width=0.31\linewidth]{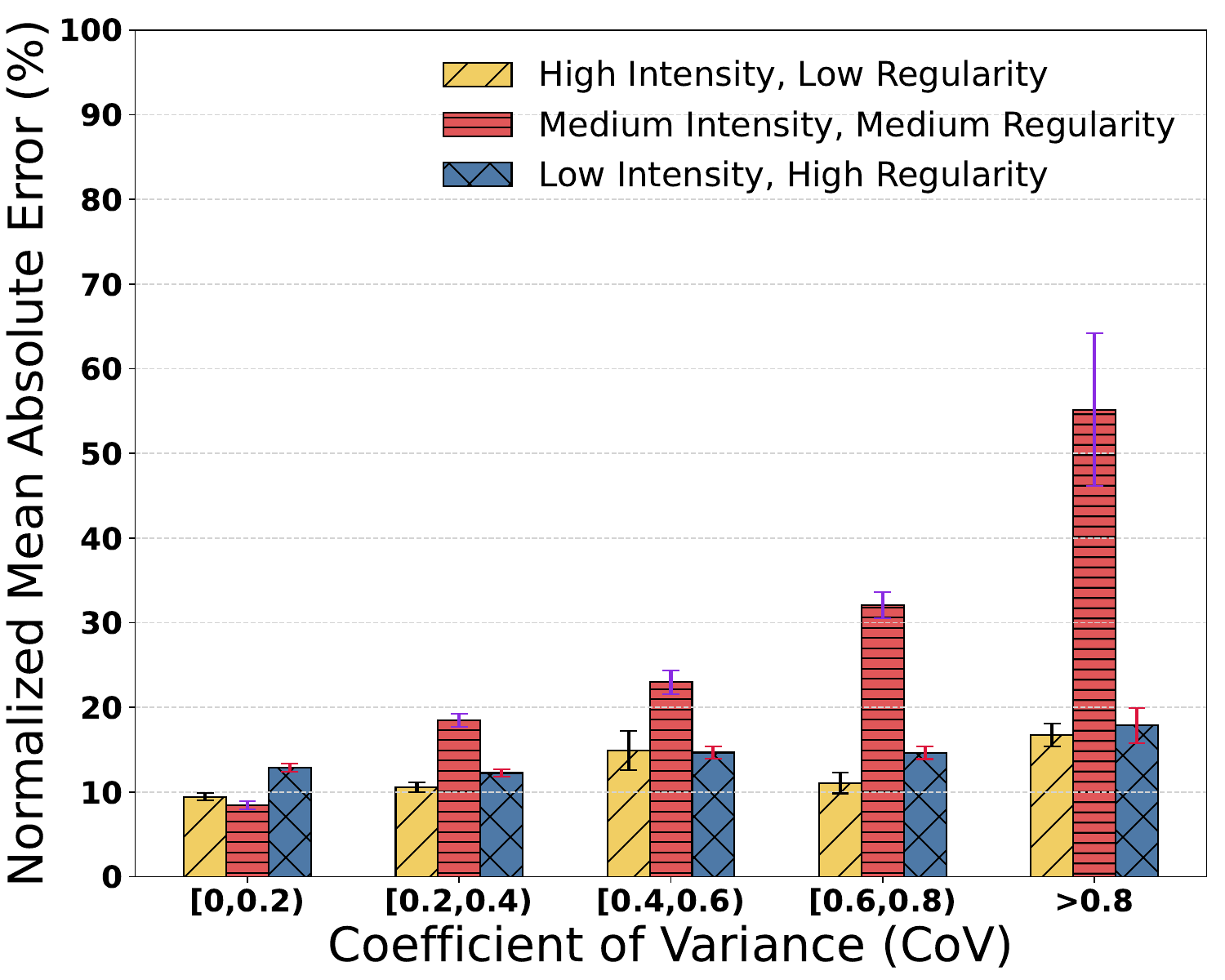} &
\includegraphics[width=0.31\linewidth]{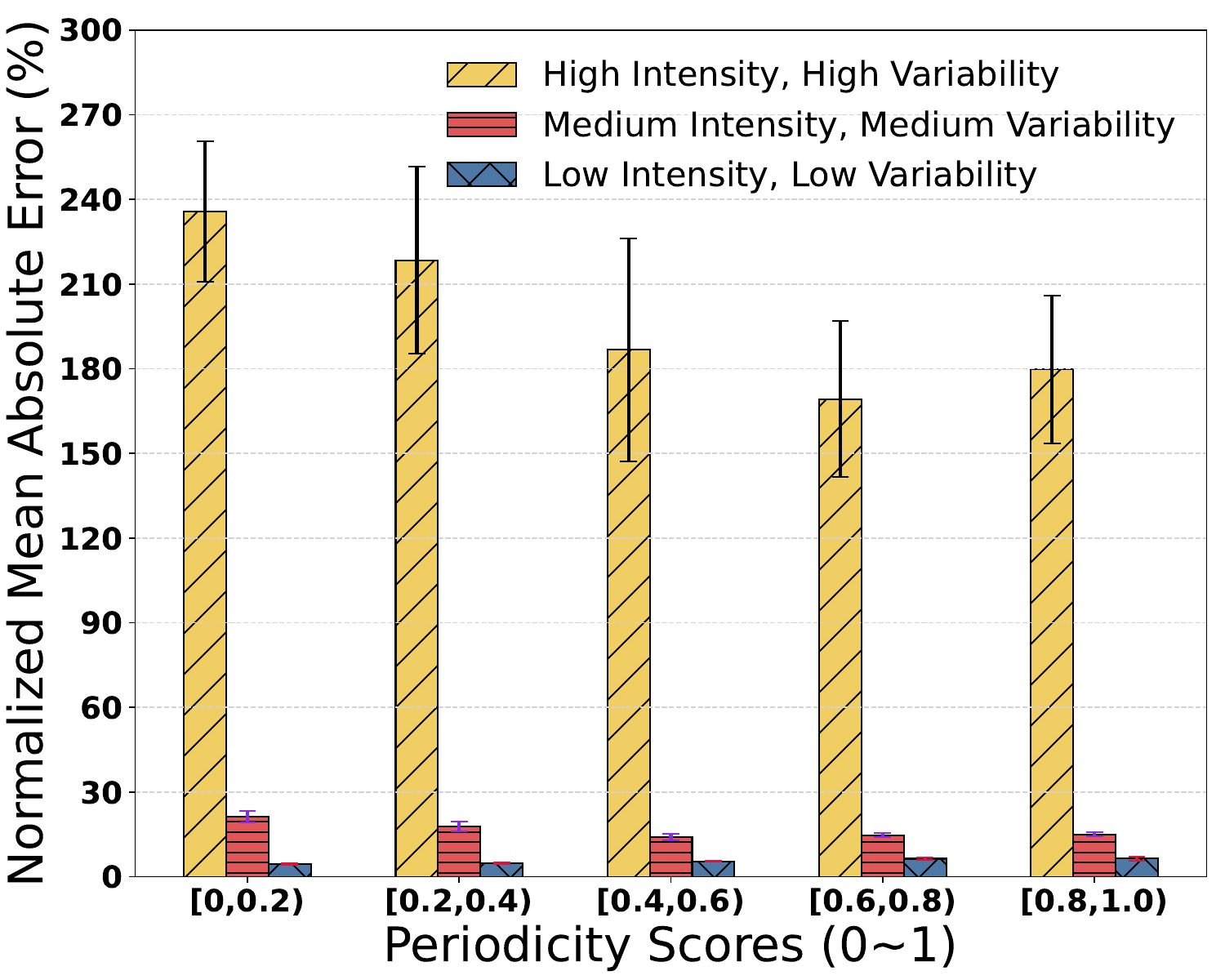} &
\includegraphics[width=0.31\linewidth]{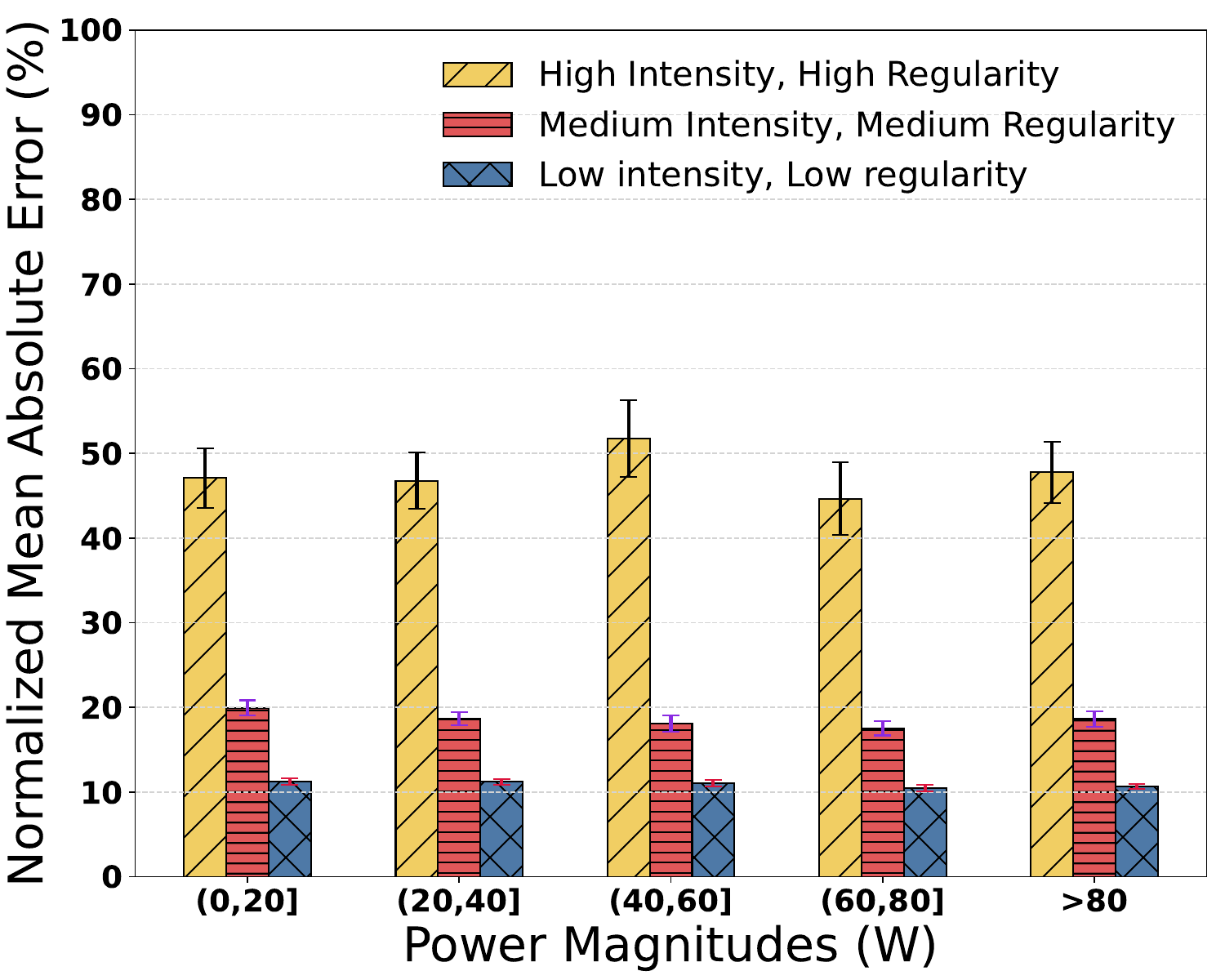} \\
(a) Effect of Variability & (b) Effect of Regularity & (c) Effect of Intensity \\
\end{tabular}
\vspace{-0.3cm}
\caption{\textbf{\emph{Effect of job characteristics: (a) effect of variability, quantified as CoV, when both regularity and intensity are controlled, (a) effect of regularity, quantified as periodicity score, when both variability and intensity are controlled, and (c) (a) effect of intensity, quantified as average power, when both 
variability and regularity are controlled. Each bar represents average across 50 jobs spread across 10 servers and error bars show the 90th percentile confidence interval across jobs. In total, each subfigure shows the disaggregation accuracy for 750 distinct jobs. }}}
\label{fig:characteristics-effect}
\vspace{-0.5cm}
\end{figure*}

\subsection{Effect of Job Characteristics}
\label{sec:job-charac}
As discussed in \S\ref{sec:workload}, a job characteristics impact  disaggregation accuracy. We next conduct experiments that decouple the effect of different job characteristics on disaggregation accuracy. In particular, we evaluate the impact of \emph{variability}, \emph{regularity}, and \emph{intensity}. To do so, we sample jobs from the Azure trace with desired characteristics and synthesize servers with desired co-location of jobs. For example, for the left-most bar in Figure~\ref{fig:characteristics-effect}(a), we select 50 jobs that have CoV between 0 and 0.2, periodicity score of less than 0.2, and magnitude of greater than 100W. We next split the 50 jobs into 10 servers each hosting 5 jobs. We then disaggregate the power of all individual jobs at once and report the average values as well as the confidence interval.  We describe the choice of jobs and their co-location settings when evaluating each factor. 

\vspace{0.1cm}
\emph{Effect of Variability.}
Figure~\ref{fig:characteristics-effect}(a) shows the disaggregation error (in NMAE) on the $y$-axis and the coefficient of variation (CoV) on the $x$-axis when the two other variables are controlled. The graph shows that, as the coefficient of variation increases, the error increases. The effect of CoV is less prominent for both low and high intensity settings since, at low and high power consumption, variability is bounded by the lower limit of 0 and the higher limit of the server's maximum power, respectively. At medium intensity, an increase in CoV results in a significant increase in variability and, thus, power disaggregation error increases.

\vspace{0.1cm}
\emph{Effect of Regularity.} Figure~\ref{fig:characteristics-effect}(b) shows the disaggregation error (in NMAE) on the $y$-axis and periodicity score (CoV) on the $x$-axis when the variability and intensity are controlled. As the periodicity score increases, we observe a downward trend in the disaggregation error, which is expected as more regular jobs are easier to disaggregate. However, we observe a very high error for the left most bar, where we have high variability (high CoV) and high intensity (high average power). This happens because periodicity is the strongest factor that affects the disaggregation accuracy. As high variability combines with high intensity, it is challenging for our disaggregator to infer the power consumption of 5 jobs that have random and high power usage.

\vspace{0.1cm}
\emph{Effect of Intensity.} Figure~\ref{fig:characteristics-effect}(c) shows the disaggregation error (in NMAE) on $y$-axis and intensity (power usage magnitude) on then $x$-axis when the variability and regularity are controlled. The effect of magnitude is only visible at the medium and low variability settings, as at high variability (yellow bar) the effect of variability dominates and results in high error with a slightly higher error at the medium magnitudes.

\noindent {\bf Key Point.} \emph{The results above show that disaggregation accuracy is a function of a job's variability, regularity, and intensity. In general, variability tends to be the dominant metric in dictating disaggregation accuracy with regularity being the next most important metric followed by intensity.}

\begin{figure}[t]
\centering
\includegraphics[width=\linewidth]{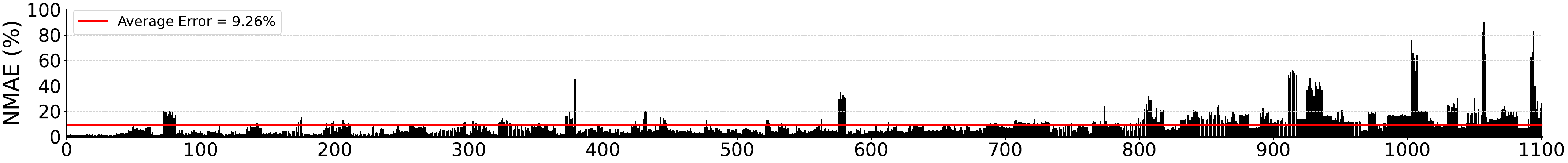}
\vspace{-0.6cm}
\caption{\textbf{\emph{Normalized Mean Absolute Error (NMAE) in disaggregating a job's power consumption on the $y$-axis for 1,100 servers from the Google trace on the $x$-axis. The average error across all the servers is 9.26\%. Each server runs 40 jobs on average. Servers are sorted in the order of increasing Coefficient of Variation (CoV) for the disaggregated job from 0.01 (left most) to 3.75 (right most).}}}
\vspace{-0.25cm}
\label{fig:google_trace_nmae}
\end{figure}

\begin{figure}[t]
\centering
\includegraphics[width=\linewidth]{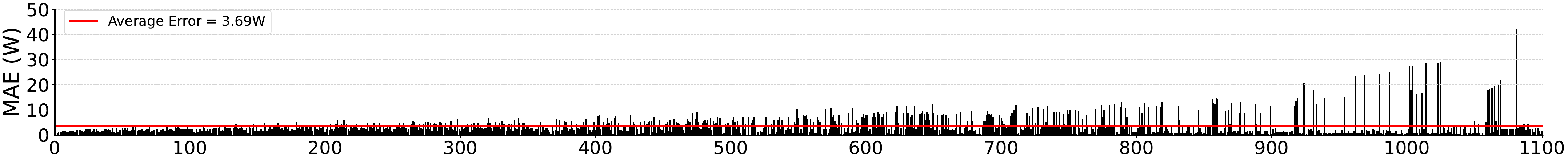}
\vspace{-0.6cm}
\caption{\textbf{\emph{Mean Absolute Error (MAE) in disaggregating a job's power consumption on the $y$-axis for 1,100 servers from Figure~\ref{fig:google_trace_nmae} on the $x$-axis. The average error across all the servers is 3.69W}.}}
\vspace{-0.5cm}
\label{fig:google_trace_mae}
\end{figure}

\subsection{Large-scale Job-level Disaggregation}
\label{sec:production}


In the previous section, we looked at the individual characteristics of the jobs where we synthesized servers with five jobs on each server, which allowed for controlled experiments. However, actual production environments have a larger number of jobs on each server and may not always place similar jobs on the same server to avoid resource contention. To evaluate the performance of \systemName in real-world settings, we use a power consumption trace based on the Google Cloud trace, which provides the actual co-location information, i.e., which jobs run on the same physical servers. We randomly selected 1,100 servers from the trace, where each server hosts 40 other jobs on average. On each server, we select one job to disaggregate, at a time, while treating others as a background jobs. As our disaggregator takes $<$1ms to disaggregate a single job for a single timestep, our method can scale to a large number of jobs.

Figure~\ref{fig:google_trace_nmae} shows the error in disaggregating power consumption of a given job on 1,100 different servers that differ in the number and characteristics of the jobs they run. We have ordered the servers by the Coefficient of Variation (CoV) for the disaggregated job from low (left) to high (right). 
We make two key observations from this experiment. First, most of the jobs (760 out of 1,100 or $\sim$69\%) have a very low error of 10\% or less, and a very small number of jobs (86 out of 1,100 or $\sim$7.81\%) have a higher than 20\% error. 
The worst-performing job has an NMAE of 90\%, but less than 3W of mean absolute error (MAE). This shows that \systemName is highly accurate in disaggregating the power consumption of jobs even in the presence of a large number of jobs on the server in practical settings. The average error is 9.20\%, which is very small considering the variations across servers and jobs. 
Second, overall, the value of NMAE increases as the CoV increases, indicating the poor disaggregation accuracy for jobs with variability in their power consumption. However, the trend is not smooth as other factors, such as the regularity and the intensity of the power consumption for a job, also affect the power disaggregation accuracy.

Figure~\ref{fig:google_trace_mae} shows the mean absolute error in disaggregating power consumption of a given job for the same set of servers as in Figure~\ref{fig:google_trace_nmae}. Most jobs (1,034 out of 1,100) have less than 10W of error. This leads to a very small average MAE of 3.69W. Even the worst-performing job has an MAE value of 42W, which is around 20\% of the maximum server power in our experimental setup.

\noindent {\bf Key Point.} \emph{Disaggregation accuracy is high for the vast majority of jobs in production due to their low variability and high regularity.}

\subsection{Scalability, Robustness, and Generalization}
\label{sec:scale-robust-general}
In this section, we evaluate \systemName's ability to scale to a large number of jobs, robustness to the number of samples used for training, and generalization in using a model trained for a given job to disaggregate another job with similar characteristics but in total different environment. 

\begin{figure}[t]
\centering
\begin{tabular}{ccc}
\includegraphics[width=0.3\linewidth]{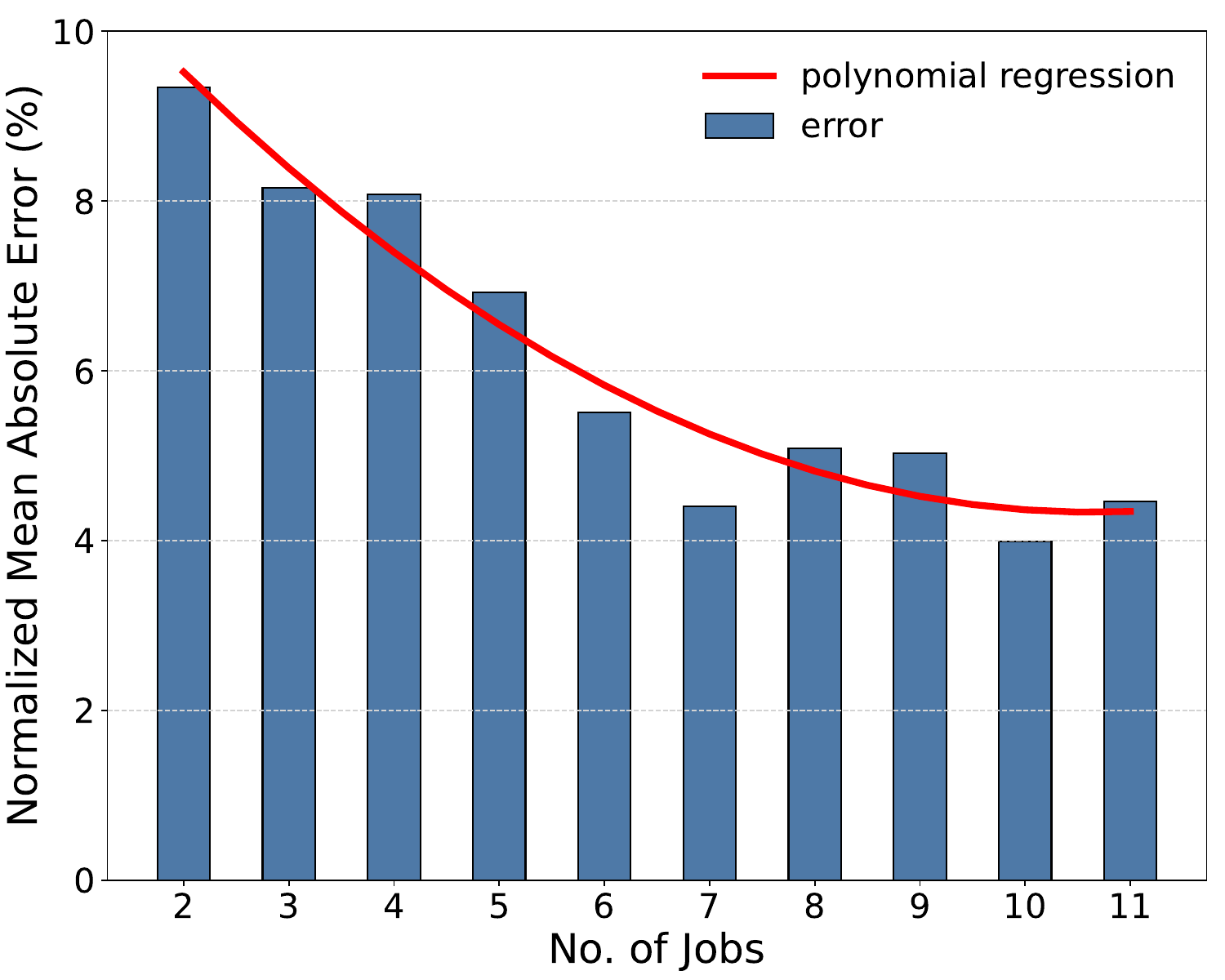} &
\includegraphics[width=0.3\linewidth]{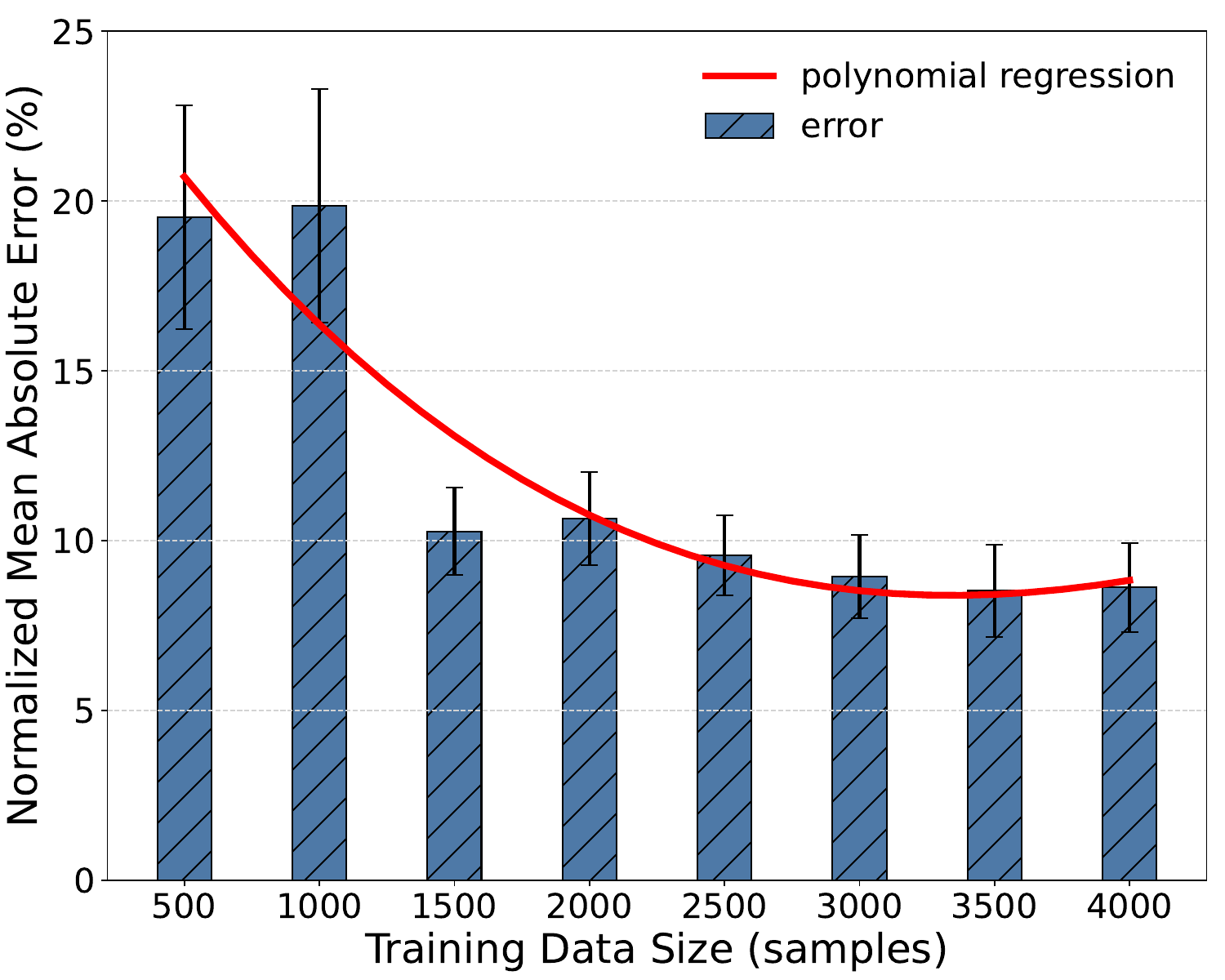} &
\includegraphics[width=0.3\linewidth]{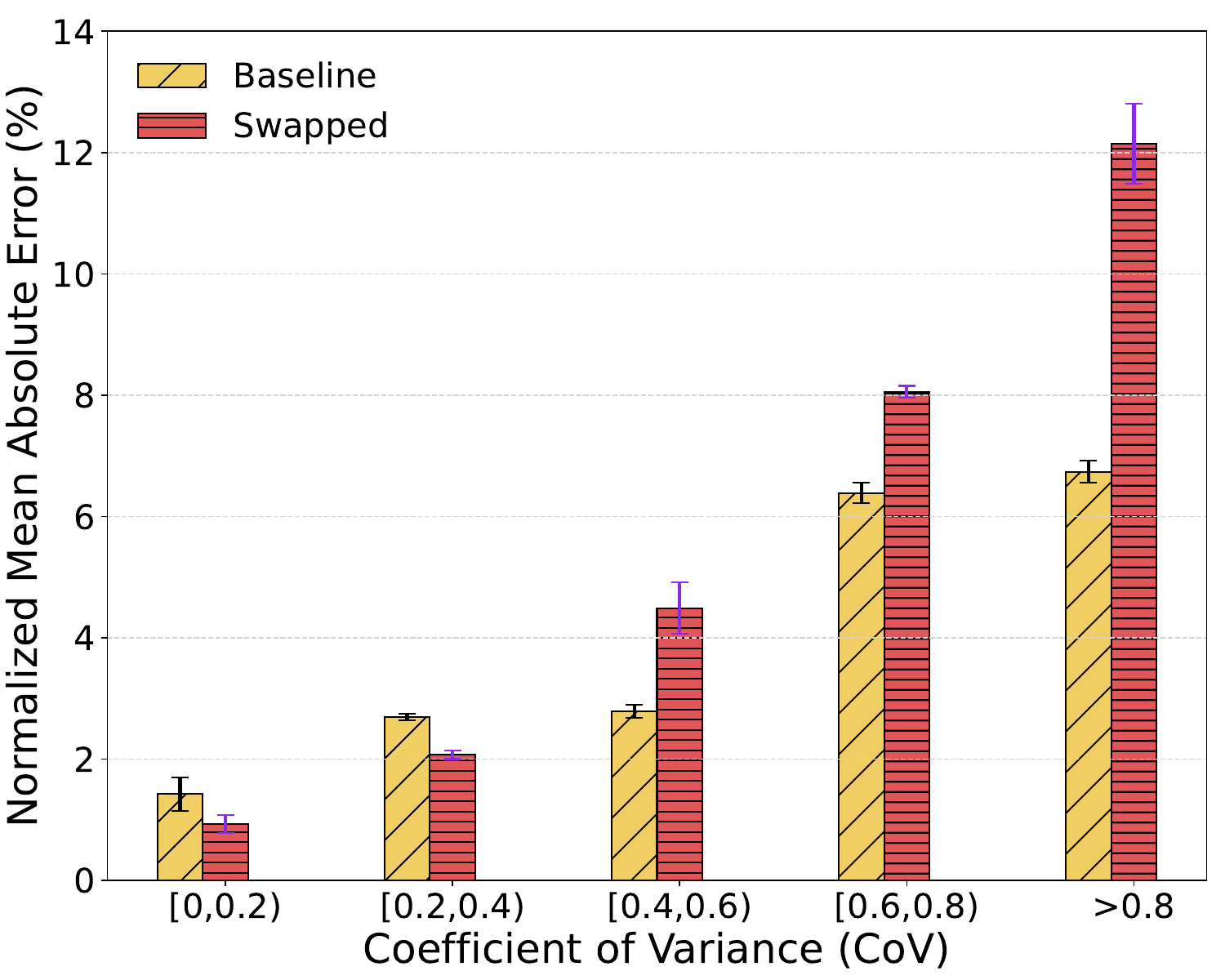} \\
(a) Scalability & (c) Robustness & (c) Generalization \\
\end{tabular}
\vspace{-0.2cm}
\caption{\textbf{\emph{\systemName Performance: (a) error in disaggregating a single job as the number of background jobs increases, (b) the effect of size of training data in number of samples used for training, and (c) generalization to disaggregating similar jobs on different servers. In (c), baseline represents the error when using the model to disaggregate  the same job that was used for training and the swapped represents the error in using the model to disaggregate a similar job on another server. Each bar in generalization results represents the average across 20 experiments.}}}
\vspace{-0.5cm}
\label{fig:scale_robust_gen}
\end{figure}

\vspace{0.1cm}
\noindent
\textbf{Scalability.}
Figure~\ref{fig:scale_robust_gen}(a) shows the average power disaggregation error as the number of jobs on the server increase. Remember, in this experiment, we disaggregate a single job against the presence of a varying number of background jobs. Each bar represents an average across 10 experiments. The decreasing trend of disaggregation error with increasing number of jobs is the result of statistical multiplexing of power usage from background jobs. When the number of jobs on a server is small, the background jobs show significant variation in their usage making the disaggregation of the desired job harder. As the number of jobs on the server increase, the variability of the background jobs decreases due to statistical multiplexing and the aggregate of background jobs becomes easier to separate from the desired job. However, it must be noted that the model used for different number of jobs in the background changes. \systemName needs to train multiple models with different number of background jobs and select an appropriate model for disaggregation at runtime, which creates a trade-off between the disaggregation accuracy and the training cost.   

\vspace{0.1cm}
\noindent
\textbf{Robustness.}
We next examine how the length of the training period for each job's model affects the power disaggregation accuracy.  Figure~\ref{fig:scale_robust_gen}(b) shows the length of the training period for the job's model (ranging from 500 samples to over 4,000 samples) on the $x$-axis and the average NMAE on the $y$-axis across all the jobs. In this case, we have on average 40 jobs co-located on each server as present in the Google trace and we are trying to disaggregate all the jobs one at a time. As expected, as the training period increases, the error tends to decrease. However, the reduction in disaggregation error is marginal once 1,500 samples have been used for training. In our case, each sample is collected over 5 minutes and the 500 samples roughly correspond to 2 days while 4,000 samples correspond to 16 days. Since these jobs are long running (31 days), using up to 6 days (1,500 samples) is feasible. It must also be noted that the wallclock time in days is purely a function of data collection granularity. If data is collected every minute instead of every 5 minutes, the same level of accuracy can be achieved using training data collected in one day. 

\vspace{0.1cm}
\noindent
\textbf{Generalization}
In our scalability experiments, we mentioned that we need to train a model for different number of background jobs which can be costly interms of training time and resources. However, the cost of training can be significantly reduced if we are able to use a single model for a similar set of jobs. To test the generalizability of \systemName's disaggregator, we train a model on a job with given CoV and use it to disaggregate a job with CoV in the same range but running on a different server. Furthermore, the other server does not have the same number of background jobs as the server used for training. 
Figure~\ref{fig:scale_robust_gen}(c) presents the results for our experiments where the $x$-axis is the coefficient of variation and $y$-axis is the average NMAE. 
The left bar (yellow, slanted pattern) represents the accuracy of the trained model on the same job while the right bar (red, horizontal pattern) represents the accuracy when the model is used on a different server to disaggregate a similar job but with different number and characteristics for the background jobs. The overall results show a high disaggregation accuracy that degrades with the increase in CoV. This indicated that the variability of the power usage trace is a stronger factor in determining the disaggregation accuracy than any other factor, even generalization. 

\noindent {\bf Key Point.}  \emph{Our experimental results show that \systemName i) scales well as more jobs run on each server, ii) is robust as the amount of training data decreases, and iii) enables the use of generalized models trained similar applications with medium-to-low CoVs at similar accuracy.}

\subsection{Production Experiments}

\begin{figure*}[t]
\centering
\begin{tabular}{ccc}
 \includegraphics[width=0.32\linewidth]{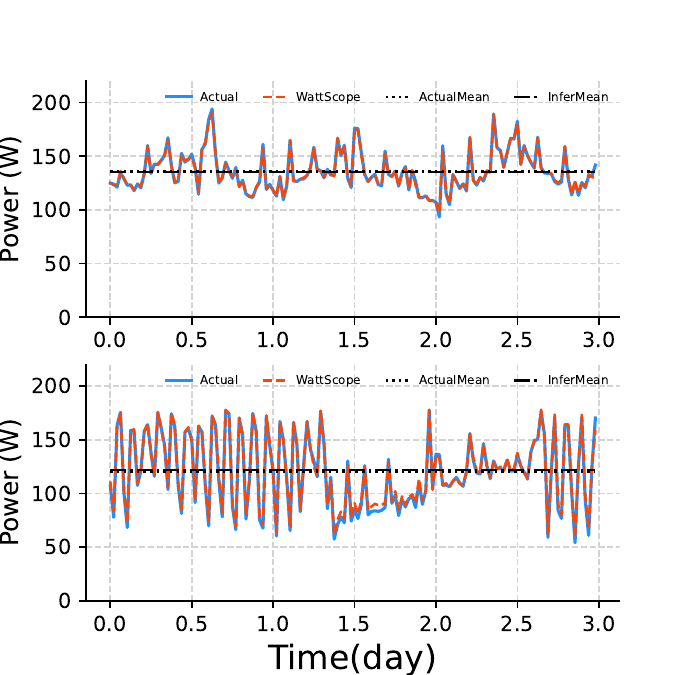} &
\includegraphics[width=0.31\linewidth]{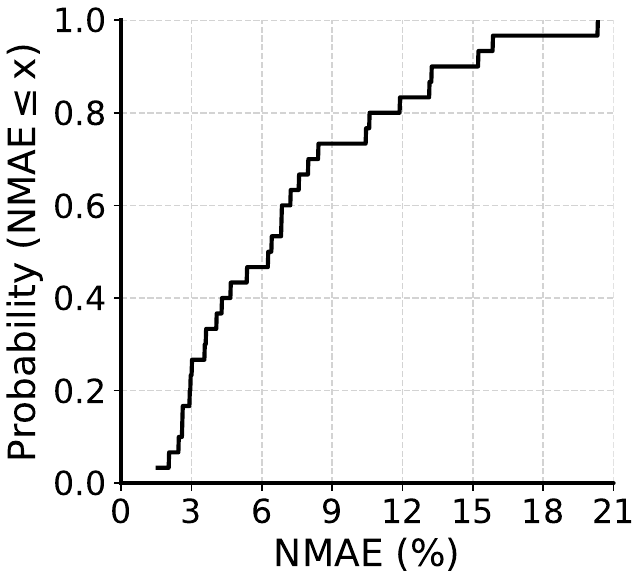} &
\includegraphics[width=0.31\linewidth]{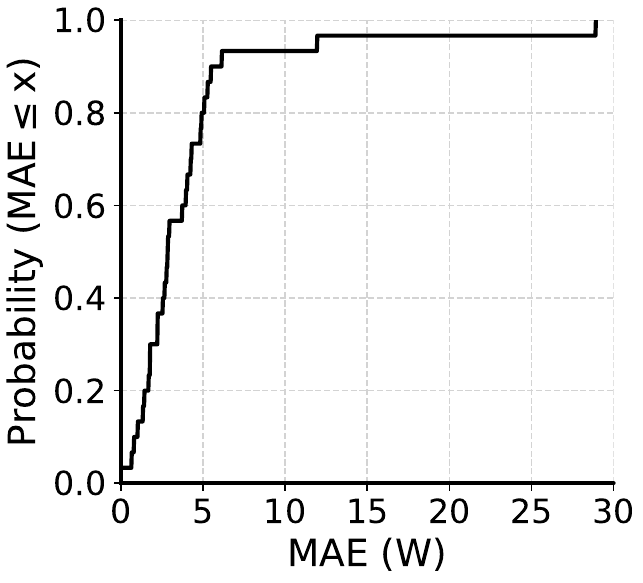} \\
(a) Illustrative Examples & (b) CDF of NMAE & (c) CDF of MAE \\
\end{tabular}
\vspace{-0.3cm}
\caption{\textbf{\emph{\systemName performance in production: (a) time-series of actual and inferred (disaggregated) power usage for the two jobs with the best NMAE (top) and the worst NMAE (bottom) from our production experiments consisting of 30 VMs, (b) NMAE distribution (7.12\% average), and (c) MAE distribution for the jobs (4.16W average). }}}
\label{fig:real-experiments}
\vspace{-0.5cm}
\end{figure*}

In this section, we evaluate \systemName's performance in disaggregating power consumption for jobs that run in a physical conventional datacenter cluster. Our cluster comprises 40 Dell PowerEdge R430s with Intel Xeon processors with 16 cores and 64GB memory. We randomly sample 30 servers from the Google Trace and replay all of the jobs on our servers using \texttt{stress-ng}~\cite{stress-ng} for 3 weeks. To get the ground-truth power consumption for a given job, we run it in isolation and record its power consumption. 

Similar to our analysis in \Section\ref{sec:qual-results}, Figure~\ref{fig:real-experiments}(a) presents the time series of actual and inferred (disaggregated) power usage for the two jobs with the best NMAE (top) and the worst NMAE (bottom) from our production experiments consisting of 30 VMs, along with the actual and inferred average power. The graph illustrates that \systemName's disaggregated power matches well with the actual power consumption observed for the job for both NMAE values. For the worst NMAE scenario, the disaggregated power deviates from the actual power consumption but closely matches the trend. Figure~\ref{fig:real-experiments}(b) and Figure~\ref{fig:real-experiments}(c) present the distribution of errors using NMAE and MAE metrics, respectively. Similar to our large-scale evaluation results, more than 70\% of the jobs have a less than 10\% NMAE, and more than 90\% of the jobs have less than 5W MAE. 

Our results demonstrate that \systemName demonstrates good performance on real power consumption traces from jobs running in conventional datacenters and can be deployed in practice.

\vspace{-0.3cm}
\section{Conclusion}
\vspace{-0.15cm}
We design a model-based system \systemName for non-intrusively estimating the power consumption of individual applications using external measurements of a server’s aggregate power usage and without requiring direct access to the server’s operating system or applications. \systemName is widely applicable in datacenters, which typically meter individual servers for management and billing. \systemName addresses key problems with traditional application-level power monitoring techniques, which are \textbf{intrusive}: require running privileged software to monitor fine-grained resource utilization and hardware support that is not always available. Our key insight (\Section\ref{sec:workload}) is that, based on an analysis of production traces, the power characteristics of datacenter workloads, e.g., low variability, low magnitude, and high periodicity, are highly amenable to disaggregation of a  server's total power consumption into application-specific values.  
We present \systemName for disaggregating server- and rack-level power meter measurements, that are already available in data centers, to server- and job-level power information, respectively. We extensively evaluate \systemName's accuracy on a production workload and show that it yields high accuracy, e.g., often $<$$\sim$10\% normalized mean absolute error.

Our key insight that enables accurate disaggregation is the generally low variability and high regularity of production applications in industry traces, as shown in \Section\ref{sec:workload}.  This insight is more broadly applicable to general scheduling and resource problems in datacenters, including placing jobs and overcommitting resources.  In the future, we plan to explore other implications of this insight. We also plan to explore methods for improving model selection by inferring an application's runtime characteristics, in terms of variability, regularity, and intensity, from its meta-data, such as the characteristics and constraints in its resource request.

\noindent {\bf Acknowledgements.} This research is supported by NSF grants 2213636, 2136199, 2106299, 2102963, 2105494, 2021693, 2020888, 2045641, as well as VMware.

\bibliographystyle{elsarticle-num-names}

\begin{thebibliography}{39}
  \expandafter\ifx\csname natexlab\endcsname\relax\def\natexlab#1{#1}\fi
  \providecommand{\url}[1]{\texttt{#1}}
  \providecommand{\href}[2]{#2}
  \providecommand{\path}[1]{#1}
  \providecommand{\DOIprefix}{doi:}
  \providecommand{\ArXivprefix}{arXiv:}
  \providecommand{\URLprefix}{URL: }
  \providecommand{\Pubmedprefix}{pmid:}
  \providecommand{\doi}[1]{\href{http://dx.doi.org/#1}{\path{#1}}}
  \providecommand{\Pubmed}[1]{\href{pmid:#1}{\path{#1}}}
  \providecommand{\bibinfo}[2]{#2}
  \ifx\xfnm\relax \def\xfnm[#1]{\unskip,\space#1}\fi
  \bibitem[{Masanet et~al.(2020)Masanet, Shehabi, Lei, Smith, and
    Koomey}]{masanet}
  \bibinfo{author}{E.~Masanet}, \bibinfo{author}{A.~Shehabi},
    \bibinfo{author}{N.~Lei}, \bibinfo{author}{S.~Smith},
    \bibinfo{author}{J.~Koomey},
  \newblock \bibinfo{title}{Recalibrating {G}lobal {D}ata {C}enter {E}nergy-use
    {E}stimates},
  \newblock \bibinfo{journal}{Science} \bibinfo{volume}{367}
    (\bibinfo{year}{2020}) \bibinfo{pages}{984--986}.
  \bibitem[{Group(2020)}]{srgresearch20}
  \bibinfo{author}{S.~R. Group}, \bibinfo{title}{Hyperscale Data Center Count
    Reaches 541 in Mid-2020; Another 176 in the Pipeline},
    \bibinfo{type}{Technical Report}, Synergy Research Group, Reno, NV (United
    States), \bibinfo{year}{2020}. \URLprefix
    \url{https://www.srgresearch.com/articles/hyperscale-data-center-count-reaches-541-mid-2020-another-176-pipeline}.
  \bibitem[{bit(2022)}]{bitcoin-energy}
  \bibinfo{title}{Digiconomist, {B}itcoin {E}nergy {C}onsumption {I}ndex},
    \bibinfo{howpublished}{\url{https://digiconomist.net/bitcoin-energy-consumption}},
    \bibinfo{year}{2022}.
  \bibitem[{Amodei et~al.(2018)Amodei, Hernandez, Sastry, Clark, Brockman, and
    Sutskever}]{ml-compute-demand}
  \bibinfo{author}{D.~Amodei}, \bibinfo{author}{D.~Hernandez},
    \bibinfo{author}{G.~Sastry}, \bibinfo{author}{J.~Clark},
    \bibinfo{author}{G.~Brockman}, \bibinfo{author}{I.~Sutskever},
    \bibinfo{title}{A{I} and {C}ompute},
    \bibinfo{howpublished}{\url{https://openai.com/blog/ai-and-compute/}},
    \bibinfo{year}{2018}.
  \bibitem[{Korn(2022)}]{growth}
  \bibinfo{author}{B.~Korn}, \bibinfo{title}{Data {C}enter {F}rontier,
    {A}chieving {E}nergy {E}fficiency in {D}ata {C}enters},
    \bibinfo{howpublished}{\url{https://datacenterfrontier.com/achieving-energy-efficiency-in-data-centers/}},
    \bibinfo{year}{2022}.
  \bibitem[{Hintemann(2018)}]{hintemann2020efficiency}
  \bibinfo{author}{R.~Hintemann}, \bibinfo{title}{Efficiency {G}ains are {N}ot
    {E}nough: {D}ata {C}enter {E}nergy {C}onsumption {C}ontinues to {R}ise
    {S}ignificantly}, \bibinfo{type}{Technical Report}, Borderstep Institute for
    Innovation and Sustainability, \bibinfo{year}{2018}.
  \bibitem[{Andrae(2019)}]{andrae2019projecting}
  \bibinfo{author}{A.~S. Andrae},
  \newblock \bibinfo{title}{Projecting the {C}hiaroscuro of the {E}lectricity
    {U}se of {C}ommunication and {C}omputing from 2018 to 2030},
  \newblock \bibinfo{journal}{Preprint}  (\bibinfo{year}{2019}).
  \bibitem[{Andrae and Edler(2015)}]{andrae2015global}
  \bibinfo{author}{A.~S. Andrae}, \bibinfo{author}{T.~Edler},
  \newblock \bibinfo{title}{On {G}lobal {E}lectricity {U}sage of {C}ommunication
    {T}echnology: {T}rends to 2030},
  \newblock \bibinfo{journal}{Challenges}  (\bibinfo{year}{2015}).
  \bibitem[{Belkhir and Elmeligi(2018)}]{belkhir2018assessing}
  \bibinfo{author}{L.~Belkhir}, \bibinfo{author}{A.~Elmeligi},
  \newblock \bibinfo{title}{Assessing {I}{C}{T} {G}lobal {E}missions {F}ootprint:
    {T}rends to 2040 \& {R}ecommendations},
  \newblock \bibinfo{journal}{Journal of Cleaner Production}
    (\bibinfo{year}{2018}).
  \bibitem[{goo(2022)}]{google-pue}
  \bibinfo{title}{Google {D}ata {C}enters: {E}fficiency},
    \bibinfo{howpublished}{\url{http://google.com/about/datacenters/efficiency/}},
    \bibinfo{year}{2022}.
  \bibitem[{ama(2019)}]{amazon-carbon-neutral}
  \bibinfo{title}{Reuters, {A}mazon {V}ows to be {C}arbon {N}eutral by 2040,
    buying 100,000 {E}lectric {V}ans},
    \bibinfo{howpublished}{\url{https://www.reuters.com/article/us-amazon-environment/amazon-vows-to-be-carbon-neutral-by-2040-buying-100000-electric-vans-idUSKBN1W41ZV}},
    \bibinfo{year}{2019}.
  \bibitem[{O'Sullivan(2020)}]{facebook-carbon-neutral}
  \bibinfo{author}{K.~O'Sullivan}, \bibinfo{title}{The {I}rish {T}imes,
    {F}acebook {C}ommits to {N}et-{Z}ero {C}arbon {E}missions by 2030},
    \bibinfo{howpublished}{\url{https://www.irishtimes.com/news/environment/facebook-commits-to-net-zero-carbon-emissions-by-2030-1.4354701}},
    \bibinfo{year}{2020}.
  \bibitem[{Acutt(2018)}]{vmware-carbon}
  \bibinfo{author}{N.~Acutt}, \bibinfo{title}{{R}adius: Stories at the {E}dge,
    {A}chieving {C}arbon {N}eutrality},
    \bibinfo{howpublished}{\url{https://www.vmware.com/radius/achieving-carbon-neutrality/}},
    \bibinfo{year}{2018}.
  \bibitem[{Etherington(2020)}]{google-carbon-free}
  \bibinfo{author}{D.~Etherington}, \bibinfo{title}{Tech{C}runch, {G}oogle
    {C}laims {N}et {Z}ero {C}arbon {F}ootprint over its {E}ntire {L}ifetime,
    {A}ims to only use {C}arbon-{F}ree {E}nergy by 2030},
    \bibinfo{howpublished}{\url{https://techcrunch.com/2020/09/14/google-claims-net-zero-carbon-footprint-over-its-entire-lifetime-aims-to-only-use-carbon-free-energy-by/-2030/}},
    \bibinfo{year}{2020}.
  \bibitem[{Smith(2020)}]{microsoft-carbon-negative}
  \bibinfo{author}{B.~Smith}, \bibinfo{title}{Official {M}icrosoft {B}log,
    {M}icrosoft will be {C}arbon {N}egative by 2030},
    \bibinfo{howpublished}{\url{https://blogs.microsoft.com/blog/2020/01/16/microsoft-will-be-carbon-negative-by-2030/}},
    \bibinfo{year}{2020}.
  \bibitem[{Pandruvada(2014)}]{rapl}
  \bibinfo{author}{S.~Pandruvada}, \bibinfo{title}{Running {A}verage {P}ower
    {L}imit},
    \bibinfo{howpublished}{\url{https://01.org/blogs/2014/running-average-power-limit-\%E2\%80\%93-rapl}},
    \bibinfo{year}{2014}.
  \bibitem[{Colmant et~al.(2015)Colmant, Kurpicz, Huertas, Rouvoy, and
    Felber}]{powerapi}
  \bibinfo{author}{M.~Colmant}, \bibinfo{author}{M.~Kurpicz},
    \bibinfo{author}{L.~Huertas}, \bibinfo{author}{R.~Rouvoy},
    \bibinfo{author}{P.~Felber},
  \newblock \bibinfo{title}{Process-level {P}ower {E}stimation in {V}{M}-based
    {S}ystems},
  \newblock in: \bibinfo{booktitle}{ACM European Conference on Computer Systems
    (EuroSys)}, \bibinfo{year}{2015}.
  \bibitem[{Colmant et~al.(2017)Colmant, Felber, Rouvoy, and
    Seinturier}]{powerapi2}
  \bibinfo{author}{M.~Colmant}, \bibinfo{author}{P.~Felber},
    \bibinfo{author}{R.~Rouvoy}, \bibinfo{author}{L.~Seinturier},
  \newblock \bibinfo{title}{Watts{K}it: Software-defined {P}ower {M}onitoring of
    {D}istributed {S}ystems},
  \newblock in: \bibinfo{booktitle}{IEEE/ACM International Symposium on Cluster,
    Cloud, and Internet Computing (CCGrid)}, \bibinfo{year}{2017}.
  \bibitem[{Bourdon et~al.(2013)Bourdon, Noureddine, Rouvoy, and
    Seinturier}]{powerapi3}
  \bibinfo{author}{A.~Bourdon}, \bibinfo{author}{A.~Noureddine},
    \bibinfo{author}{R.~Rouvoy}, \bibinfo{author}{L.~Seinturier},
  \newblock \bibinfo{title}{Power{A}{P}{I}: A {S}oftware {L}ibrary to {M}onitor
    the {E}nergy {C}onsumed at the {P}rocess-level},
  \newblock \bibinfo{journal}{ERCIM News, Special Theme: Smart Energy Systems}
    (\bibinfo{year}{2013}) \bibinfo{pages}{43--44}.
  \bibitem[{Batra et~al.(2019)Batra, Kukunuri, Pandey, Malakar, Kumar,
    Krystalakos, Zhong, Meira, and Parson}]{nilmtk-contrib}
  \bibinfo{author}{N.~Batra}, \bibinfo{author}{R.~Kukunuri},
    \bibinfo{author}{A.~Pandey}, \bibinfo{author}{R.~Malakar},
    \bibinfo{author}{R.~Kumar}, \bibinfo{author}{O.~Krystalakos},
    \bibinfo{author}{M.~Zhong}, \bibinfo{author}{P.~Meira},
    \bibinfo{author}{O.~Parson},
  \newblock \bibinfo{title}{Towards {R}eproducible {S}tate-of-the-{A}rt {E}nergy
    {D}isaggregation},
  \newblock in: \bibinfo{booktitle}{ACM International Conference on Systems for
    Energy-Efficient Buildings, Cities, and Transportation (BuildSys)},
    \bibinfo{year}{2019}.
  \bibitem[{sec(2022)}]{sec-rule}
  \bibinfo{title}{Press {R}elease, {S}{E}{C} {P}roposes {R}ules to {E}nhance and
    {S}tandardize {C}limate-{R}elated {D}isclosures for {I}nvestors},
    \bibinfo{howpublished}{\url{https://www.sec.gov/news/press-release/2022-46}},
    \bibinfo{year}{2022}.
  \bibitem[{Hanafy et~al.(2023)Hanafy, Bostandoost, Bashir, Irwin, Hajiesmaili,
    and Shenoy}]{war}
  \bibinfo{author}{W.~A. Hanafy}, \bibinfo{author}{R.~Bostandoost},
    \bibinfo{author}{N.~Bashir}, \bibinfo{author}{D.~Irwin},
    \bibinfo{author}{M.~Hajiesmaili}, \bibinfo{author}{P.~Shenoy},
  \newblock \bibinfo{title}{The {W}ar of the {E}fficiencies: Understanding the
    {T}ension between {C}arbon and {E}nergy {O}ptimization},
  \newblock in: \bibinfo{booktitle}{Proceedings of the Workshop on Sustainable
    Computer Systems (HotCarbon)}, \bibinfo{year}{2023}.
  \bibitem[{ipm(2022)}]{ipmi}
  \bibinfo{title}{Intelligent platform management interface specification},
    \bibinfo{howpublished}{\url{https://www.intel.com/content/www/us/en/products/docs/servers/ipmi/ipmi-second-gen-interface-spec-v2-rev1-1.html}},
    \bibinfo{year}{2022}.
  \bibitem[{red(2022)}]{redfish}
  \bibinfo{title}{D{M}{T}{F}: Redfish {D}evelopers {H}ub},
    \bibinfo{howpublished}{\url{https://redfish.dmtf.org/}},
    \bibinfo{year}{2022}.
  \bibitem[{Bourdon et~al.(2013)Bourdon, Noureddine, Rouvoy, and
    Seinturier}]{power-api}
  \bibinfo{author}{A.~Bourdon}, \bibinfo{author}{A.~Noureddine},
    \bibinfo{author}{R.~Rouvoy}, \bibinfo{author}{L.~Seinturier},
  \newblock \bibinfo{title}{{Power{API}: {A} {S}oftware {L}ibrary to {M}onitor
    the {E}nergy {C}onsumed at the {Pr}ocess-{L}evel}},
  \newblock \bibinfo{journal}{{ERCIM News}} \bibinfo{volume}{92}
    (\bibinfo{year}{2013}) \bibinfo{pages}{43--44}. \URLprefix
    \url{https://inria.hal.science/hal-00772454}.
  \bibitem[{Cortez et~al.(2017)Cortez, Bonde, Muzio, Russinovich, Fontoura, and
    Bianchini}]{azure-trace}
  \bibinfo{author}{E.~Cortez}, \bibinfo{author}{A.~Bonde},
    \bibinfo{author}{A.~Muzio}, \bibinfo{author}{M.~Russinovich},
    \bibinfo{author}{M.~Fontoura}, \bibinfo{author}{R.~Bianchini},
  \newblock \bibinfo{title}{Resource {C}entral: Understanding and {P}redicting
    {W}orkloads for {I}mproved {R}esource {M}anagement in {L}arge {C}loud
    {P}latforms},
  \newblock in: \bibinfo{booktitle}{Proceedings of the Symposium on Operating
    Systems Principles (SOSP)}, \bibinfo{publisher}{ACM}, \bibinfo{year}{2017}.
  \bibitem[{Tirmazi et~al.(2020)Tirmazi, Barker, Deng, Haque, Qin, Hand,
    Harchol-Balter, and Wilkes}]{google-trace}
  \bibinfo{author}{M.~Tirmazi}, \bibinfo{author}{A.~Barker},
    \bibinfo{author}{N.~Deng}, \bibinfo{author}{M.~E. Haque},
    \bibinfo{author}{Z.~G. Qin}, \bibinfo{author}{S.~Hand},
    \bibinfo{author}{M.~Harchol-Balter}, \bibinfo{author}{J.~Wilkes},
  \newblock \bibinfo{title}{Borg: The {N}ext {G}eneration},
  \newblock in: \bibinfo{booktitle}{ACM European Conference on Computer Systems
    (EuroSys)}, \bibinfo{publisher}{ACM}, \bibinfo{address}{New York, NY, USA},
    \bibinfo{year}{2020}.
  \bibitem[{King(2023)}]{stress-ng}
  \bibinfo{author}{C.~I. King}, \bibinfo{title}{stress-ng},
    \bibinfo{howpublished}{\url{https://github.com/ColinIanKing/stress-ng}},
    \bibinfo{year}{2023}.
  \bibitem[{Elfeky et~al.(2005)Elfeky, Aref, and Elmagarmid}]{period-detection}
  \bibinfo{author}{M.~Elfeky}, \bibinfo{author}{W.~Aref},
    \bibinfo{author}{A.~Elmagarmid},
  \newblock \bibinfo{title}{Periodicity {D}etection in {T}ime {S}eries
    {D}atabases},
  \newblock \bibinfo{journal}{IEEE Transactions on Knowledge and Data Engineering
    (TKDE)} \bibinfo{volume}{17} (\bibinfo{year}{2005}).
  \bibitem[{Kumbhare et~al.(2021)Kumbhare, Azimi, Manousakis, Bonde, Frujeri,
    Mahalingam, Misra, Javadi, Schroeder, Fontoura, and Bianchini}]{predictions}
  \bibinfo{author}{A.~Kumbhare}, \bibinfo{author}{R.~Azimi},
    \bibinfo{author}{I.~Manousakis}, \bibinfo{author}{A.~Bonde},
    \bibinfo{author}{F.~Frujeri}, \bibinfo{author}{N.~Mahalingam},
    \bibinfo{author}{P.~Misra}, \bibinfo{author}{S.~Javadi},
    \bibinfo{author}{B.~Schroeder}, \bibinfo{author}{M.~Fontoura},
    \bibinfo{author}{R.~Bianchini},
  \newblock \bibinfo{title}{Prediction-based {P}ower {O}versubscription in
    {C}loud {P}latorms},
  \newblock in: \bibinfo{booktitle}{USENIX Annual Technical Conference (ATC)},
    \bibinfo{year}{2021}.
  \bibitem[{azu(2022)}]{azure-data-explorer}
  \bibinfo{title}{Microsoft {I}gnite, anomaly {D}etection and {F}orecasting in
    {A}zure {D}ata {E}xplorer},
    \bibinfo{howpublished}{\url{https://learn.microsoft.com/en-us/azure/data-explorer/anomaly-detection}},
    \bibinfo{year}{2022}.
  \bibitem[{per(2022)}]{periods-detect}
  \bibinfo{title}{Microsoft {I}gnite, series\_periods\_detect()},
    \bibinfo{howpublished}{\url{https://learn.microsoft.com/en-us/azure/data-explorer/kusto/query/series-periods-detectfunction}},
    \bibinfo{year}{2022}.
  \bibitem[{Barker et~al.(2014)Barker, Kalra, Irwin, and Shenoy}]{barker:jsac}
  \bibinfo{author}{S.~Barker}, \bibinfo{author}{S.~Kalra},
    \bibinfo{author}{D.~Irwin}, \bibinfo{author}{P.~Shenoy},
  \newblock \bibinfo{title}{Empirical {C}haracterization, {M}odeling, and
    {A}nalysis of {S}mart {M}eter {D}ata},
  \newblock \bibinfo{journal}{IEEE Journal on Selected Areas of Communications
    (JSAC), Smart Grid Communications Series} \bibinfo{volume}{32}
    (\bibinfo{year}{2014}) \bibinfo{pages}{1312--1327}.
  \bibitem[{Barker et~al.(2013)Barker, Kalra, Irwin, and Shenoy}]{barker:igcc13}
  \bibinfo{author}{S.~Barker}, \bibinfo{author}{S.~Kalra},
    \bibinfo{author}{D.~Irwin}, \bibinfo{author}{P.~Shenoy},
  \newblock \bibinfo{title}{Empirical {C}haracterization and {M}odeling of
    {E}lectrical {L}oads in {S}mart {H}omes},
  \newblock in: \bibinfo{booktitle}{Proceedings of the 4th IEEE International
    Green Computing Conference (IGCC)}, \bibinfo{address}{Arlington, Virginia},
    \bibinfo{year}{2013}, pp. \bibinfo{pages}{1--10}.
  \bibitem[{Verma et~al.(2015)Verma, Pedrosa, Korupolu, Oppenheimer, Tune, and
    Wilkes}]{borg}
  \bibinfo{author}{A.~Verma}, \bibinfo{author}{L.~Pedrosa},
    \bibinfo{author}{M.~Korupolu}, \bibinfo{author}{D.~Oppenheimer},
    \bibinfo{author}{E.~Tune}, \bibinfo{author}{J.~Wilkes},
  \newblock \bibinfo{title}{Large-scale {C}luster {M}anagement at {G}oogle with
    {B}org},
  \newblock in: \bibinfo{booktitle}{ACM European Conference on Computer Systems
    (EuroSys)}, \bibinfo{address}{Bordeaux, France}, \bibinfo{year}{2015}, pp.
    \bibinfo{pages}{1--17}.
  \bibitem[{Bashir et~al.(2021)Bashir, Deng, Rzadca, Irwin, Kodak, and
    Jnagal}]{limit-eurosys}
  \bibinfo{author}{N.~Bashir}, \bibinfo{author}{N.~Deng},
    \bibinfo{author}{K.~Rzadca}, \bibinfo{author}{D.~Irwin},
    \bibinfo{author}{S.~Kodak}, \bibinfo{author}{R.~Jnagal},
  \newblock \bibinfo{title}{Take it to the {L}imit: Prediction-{D}riven
    {R}esource {O}vercommitment in {D}atacenters},
  \newblock in: \bibinfo{booktitle}{ACM European Conference on Computer Systems
    (EuroSys)}, \bibinfo{year}{2021}.
  \bibitem[{Batra et~al.(2014)Batra, Kelly, Parson, Dutta, Knottenbelt, Rogers,
    Singh, and Srivastava}]{nilmtk}
  \bibinfo{author}{N.~Batra}, \bibinfo{author}{J.~Kelly},
    \bibinfo{author}{O.~Parson}, \bibinfo{author}{H.~Dutta},
    \bibinfo{author}{W.~Knottenbelt}, \bibinfo{author}{A.~Rogers},
    \bibinfo{author}{A.~Singh}, \bibinfo{author}{M.~Srivastava},
  \newblock \bibinfo{title}{Nilmtk: An open source toolkit for non-intrusive load
    monitoring},
  \newblock in: \bibinfo{booktitle}{ACM International Conference on Future Energy
    Systems (e-Energy)}, \bibinfo{year}{2014}.
  \bibitem[{Krystalakos et~al.(2018)Krystalakos, Nalmpantis, and
    Vrakas}]{sliding-window}
  \bibinfo{author}{O.~Krystalakos}, \bibinfo{author}{C.~Nalmpantis},
    \bibinfo{author}{D.~Vrakas},
  \newblock \bibinfo{title}{Sliding {W}indow {A}pproach for {O}nline {E}nergy
    {D}isaggregation using {A}rtificial {N}eural {N}etworks},
  \newblock in: \bibinfo{booktitle}{Hellenic Conference on Artificial
    Intelligence}, \bibinfo{year}{2018}.
  \bibitem[{Kansal et~al.(2010)Kansal, Zhao, Liu, Kothari, and
    Bhattacharya}]{VM-Power}
  \bibinfo{author}{A.~Kansal}, \bibinfo{author}{F.~Zhao},
    \bibinfo{author}{J.~Liu}, \bibinfo{author}{N.~Kothari},
    \bibinfo{author}{A.~A. Bhattacharya},
  \newblock \bibinfo{title}{Virtual {M}achine {P}ower {M}etering and
    {P}rovisioning},
  \newblock in: \bibinfo{booktitle}{ACM Symposium on Cloud Computing (SoCC)},
    \bibinfo{publisher}{ACM}, \bibinfo{address}{New York, NY, USA},
    \bibinfo{year}{2010}, p. \bibinfo{pages}{39–50}.
  
  \end{thebibliography}

\end{document}